\documentclass[twocolumn,trackchanges]{aastex62}
\usepackage{amsmath}
\usepackage[utf8]{inputenc}
\usepackage{multirow}
\usepackage{dcolumn}

\received{January 1, 2018}
\revised{January 7, 2018}
\accepted{\today}
\submitjournal{ApJ}

\shorttitle{Self-consistent solutions for line-driven winds of hot massive stars}
\shortauthors{Gormaz-Matamala et al.}
\begin{document}

\title{Self-consistent solutions for line-driven winds of hot massive stars\\The m-CAK procedure}

\correspondingauthor{Alex C. Gormaz-Matamala}
\email{alex.gormaz@postgrado.uv.cl}

\author[0000-0002-0786-7307]{Alex C. Gormaz-Matamala}
\affil{Instituto de F\'isica y Astronom\'ia, Universidad de Valpara\'iso, Av. Gran Breta\~na 1111, Casilla 5030, Valpara\'iso, Chile}
\affiliation{Centro de Astrofísica, Universidad de Valparaíso. Av. Gran Breta\~na 1111, Casilla 5030, Valpara\'io, Chile.}

\author[0000-0002-2191-8692]{M. Cur\'e}
\affil{Instituto de F\'isica y Astronom\'ia, Universidad de Valpara\'iso, Av. Gran Breta\~na 1111, Casilla 5030, Valpara\'iso, Chile}
\affiliation{Centro de Astrofísica, Universidad de Valparaíso. Av. Gran Breta\~na 1111, Casilla 5030, Valpara\'io, Chile.}

\author{L. S. Cidale$^{*}$ \thanks{Member of Carrera del Investigador Cient\'ifico de CONICET}}
%\affil{Instituto de F\'isica y Astronom\'ia, Universidad de Valpara\'iso, Av. Gran Breta\~na 1111, Casilla 5030, Valpara\'iso, Chile}
\affiliation{Departamento de Espectroscop\'ia, Facultad de Ciencias Astron\'omicas y Geof\'isicas, UNLP. Paseo del Bosque S/N, 1900 La Plata, Argentina.}
\affiliation{Instituto de Astrof\'isica de La Plata, CCT La Plata, CONICET-UNLP. Paseo del Bosque S/N, 1900 La Plata, Argentina.}
\thanks{Member of the Carrera del Investigador Cient\'{\i}fico,\\ CONICET, Argentina}

\author{R. O. J. Venero}
\affiliation{Departamento de Espectroscop\'ia, Facultad de Ciencias Astron\'omicas y Geof\'isicas, UNLP. Paseo del Bosque S/N, 1900 La Plata, Argentina.}
\affiliation{Instituto de Astrof\'isica de La Plata, CCT La Plata, CONICET-UNLP. Paseo del Bosque S/N, 1900 La Plata, Argentina.}



\begin{abstract}
	Massive stars present strong stellar that which are described by the radiation driven wind theory. Accurate mass-loss rates are necessary to properly describe the stellar evolution across the Hertzsprung--Russel Diagram.
	We present a self-consistent procedure that coupled the hydrodynamics with calculations of the line-force, giving as results the line-force parameters, the  velocity field, and the mass-loss rate. Our calculations contemplate the contribution to the line-force multiplier from more than $\sim 900,000$ atomic transitions, an NLTE radiation flux from the photosphere and a quasi-LTE approximation for the occupational numbers.
	A full set of line-force parameters for $T_\text{eff}\ge 32,000$ K, surface gravities higher than 3.4 dex for two different metallicities are presented, with their corresponding wind parameters (terminal velocities and mass-loss rates).
	The already known dependence of line-force parameters on effective temperature is enhanced by the dependence on $\log g$. The terminal velocities present a stepper scaling relation with respect to the escape velocity, this might explain the scatter values observed in the hot side of the bistability jump.
	Moreover, a comparison of self-consistent mass-loss rates with empirical values shows a good agreement. Self-consistent wind solutions are used as input in FASTWIND to calculate synthetic spectra. We show, comparing with the observed spectra for three stars, that varying the clumping factor, the synthetic spectra rapidly converge into the neighbourhood region of the solution. It is important to stress that our self-consistent procedure significantly reduces the number of free parameters needed to obtain a synthetic spectrum.
\end{abstract}



\section{Introduction}\label{sec:intro}
	The study of massive stars (i.e., stars with  $M_* >10\,M_\odot$) is a relevant topic in the framework of stellar astrophysics, because these stars exhibit some of the most extreme physical conditions, such as the hottest temperatures, the highest outflows of matter, and a complex nucleosynthesis.

	Strong outflowing stellar winds of massive stars eject high amounts of matter that contribute to the chemical enrichment of the ISM in a relatively short timescale. Moreover, it has been found that differences on a factor of two in the mass-loss rate considerably affects the final fate of a star \citep{meynet94,smith14}.
	Therefore, a better understanding about massive stars and their evolution strongly requires accurate determination of their fundamental parameters, with the mass-loss rate being the most relevant \citep[][]{kudritzki00,puls08}.
	
	\citet{lucy70} described the mechanism that drives the strong stellar winds observed in hot stars: the so-called radiation driven winds.
	According to these authors, the absorption and further reemission of photons by UV resonance lines is the wind-driven mechanism for hot stars, that produces an outward line-force.
	The foundation of the theory of radiation driven winds was later developed by \citet[][hereafter CAK theory]{cak}, who, based on the Sobolev and the point-star approximations, modelled the line-acceleration analytically in terms of the acceleration produced by electron scattering times a force multiplier factor.
	This factor represents the contribution of absorption and reemission processes depending on the optical depth only, and it was parametrised by two constant parameters through the wind, namely $k$ and $\alpha$. 

	Later, \citet{abbott82} performed a detailed calculation of these line-force parameters taking into account the contribution of a full set of atomic line transition data for elements from hydrogen to zinc.
	Due to the point-star approximation the derived hydrodynamical values for mass-loss rates were overestimated; \citet{ppk} and \citet{friend86} relaxed this approximation and considered the finite-disk shape of the star.
	With this modified theory (hereafter m-CAK), they solved the equation of motion and obtained improved theoretical results, in better agreement with the observed mass-loss rates.
	
	Due to scarce works concerning  NLTE (nonlocal thermodynamic equilibrium) calculations of the line-force parameters \citep{ppk,puls00,kudritzki02,pauldrach03,noebauer15}, it was difficult to obtain from the m-CAK hydrodynamics the velocity profiles and mass-loss rates; thus, the massive star community started to use the so-called $\beta$-law velocity profile.
	This simplified description of the velocity field is widely used as input in radiative transfer and NLTE model-atmosphere codes such as FASTWIND \citep{santolaya97,puls05} or CMFGEN \citep{hillier90,hillier98,hillier01} to calculate synthetic spectra.
	In this procedure, stellar and wind parameters (terminal velocity and mass-loss rates) are treated as free and are determined by varying them to adjust synthetic to observed line profiles.
	\citet{kudritzki00} argued that the use of $\beta$-law for the velocity field is only justified \textit{a posteriori} once the fit is achieved. There are other approaches that coupled the hydrodynamics with comoving frame radiative transfer, see, e.g. \cite{sander17} or \cite{kk10,kk17}, that do not use a $\beta$-law velocity profile.

	Calculations of line-force wind parameters coupled with  hydrodynamics are necessary to derive self-consistent velocity profiles and mass-loss rates.
	Moreover, they depend nonlinearly on the stellar parameters, chemical abundances, and atomic data via the wind-driven mechanism.
	To obtain the line-force parameters it is necessary to calculate the \textit{total acceleration} produced by the contribution of hundreds of thousands of lines participating in the absorption and reemission processes (hereafter line-acceleration, $g_\text{line}$).
	Thus, having reliable atomic data is essential to perform \textit{line-statistics} calculations.
	
	The number of contributing lines to the driven line-acceleration depends on the wind opacity and it is strongly coupled to the wind density and velocity profiles.
	To solve this highly nonlinear problem an iterative procedure is required to satisfy both: line-statistics and m-CAK hydrodynamics.

	In this work, we calculate self-consistent solutions to obtain accurate m-CAK line-force parameters $(k,\alpha,\delta)$ and wind properties of hot massive stars.
	The hydrodynamics is provided using our code \textsc{HydWind} \citep{michel04}, whereas abundances have been updated from \citet{asplund09}. 
	Final self-consistent line-force values correspond to a unique solution obtained when line-force parameters, velocity profile and mass-loss rate converged.
	Hence, we present here a new set of m-CAK self-consistent line-force parameters  for   $T_\text{eff} > 32\,000 K$ and  $\log g\ge3.4$, with the corresponding velocity profile  and mass-loss rate.
	These line-force parameters are compared with previous numerical studies. Furthermore, with these new results we calculate synthetic spectra with FASTWIND contrasting them with observations. We show that applying few times our procedure we obtain a very good fit of the observed line profile. Furthermore, we derived (i) an alternative recipe for the mass-loss rate which only depends on the stellar parameters and the abundance; (ii) the ratio $v_{\inf}/v_\text{esc}$ as given by Eq. (\ref{cinfinit}) now depends not only on the line-force parameter $\alpha$ but also on $\log g$.
		
	This paper is organised as follows: The theoretical formulation of m-CAK theory is given in Section \ref{theory}.
	Section \ref{methods} describes the methodology used, explaining the iterative procedure and how convergence is assured.
	Section \ref{results} shows results for the calculation of the line-force multiplier using the standard solution, together with a detailed analysis about under what conditions $(k,\alpha,\delta)$ can be treated as constants. In Section \ref{syntheticspectra}, we calculate synthetic spectra based on our self-consistent procedure and compare them with observations.
	A discussion about the results is given in Section \ref{discussion}.
	Finally, our conclusions are presented in Section \ref{conclusions}.

%_____THEORY_____________________________________________________________________________________
\section{Theoretical formulation}\label{theory}

	The m-CAK theory \citep{cak,friend86,ppk} describes in spherical coordinates a stationary, nonrotating, expanding atmosphere, taking into account the line-acceleration $g_\text{line}$.
	The equation of momentum and equation of continuity respectively read:
	\begin{equation}\label{momentum}
		 v\frac{d v}{dr}=-\frac{1}{\rho}\frac{dP}{dr}-\frac{GM_*(1-\Gamma_e)}{r^2}+g_\text{line}\;,
	\end{equation}
	and
	\begin{equation}
		\dot M=4\pi\rho(r)r^2 v(r)\;.
	\end{equation}
	Here, $\dot{M}$ is the mass-loss rate, $v(r)$ is the radial velocity field, $\rho(r)$ is the mass density, $P$ is the gas pressure and $M_*(1-\Gamma_e)$ corresponds to the effective stellar mass, where $\Gamma_e$ is the radiative acceleration caused by Thomson scattering in terms of gravitational acceleration.

	Introducing the following dimensionless variables: $\hat{r}=r/R_{*}$, $\hat{v}=v/a$ and $\hat{v}_{\rm{crit}}=v_{\rm{esc}}/a\sqrt{2}$, where the escape velocity is defined as $v_{\rm{esc}}^2=2\, G\, M_* (1-\Gamma_e)/R_*$. Then, the equation of motion reads:
	
	\begin{equation}\label{motion}
		\left(\hat v-\frac{1}{\hat v}\right)\frac{d\hat v}{d\hat r}=-\frac{\hat v_\text{crit}^2}{\hat r^2}+\frac{2}{\hat r}+\hat g_\text{line}\;,
	\end{equation}
	\noindent with $\hat{g}_{\rm{line}}=(R_{*}/a^{2})\, g_{\rm{line}}$.
	We have used the equation of state of an ideal gas, $P=a^2\rho$, with $a$ being the isothermal sound speed:
		\begin{equation}\label{sound_plus_vturb}
		a=\sqrt{\frac{k_BT_\text{eff}}{\mu\,m_H}}\,\,,
		\end{equation}
	with $\mu$ being the mean atomic weight.

	The line-acceleration can be defined in terms of the radiative acceleration due to electron scattering $\hat{g}_{e}= (R_{*}/a^{2})\, g_e$,  multiplied by $\mathcal M(t)=\hat{g}_\text{line}/\hat{g_e}$, called the \textit{line-force multiplier factor}. $\mathcal M(t)$ corresponds to the sum of spectral lines that contribute to drive the wind:
	\begin{equation}\label{mt_definition}
		\mathcal{M}(t)=\sum_{\text{lines}}\Delta\nu_D\,\frac{F_\nu}{F}\frac{1-e^{-\eta_{\rm{line}}\, t}}{t}\;,
	\end{equation}
	with $\Delta\nu_D$ being the line broadening due to Doppler effects. $F_\nu$ and $F$ are the monochromatic and total stellar flux, respectively, and $\eta_{\rm{line}}$ is the absorption coefficient.
	\citet{castor74} parametrised $\mathcal M(t)$  in terms of the optical depth $t$, which depends on the wind structure only:
	\begin{equation}\label{t}
		t=\sigma_e\,\rho(r)\, v_\text{th}\left(\frac{d v}{dr}\right)^{-1}\;\;,
	\end{equation}
	with $v_\text{th}$ being the mean hydrogen thermal velocity.
	
	Then, \citet{cak} proposed the following analytical expression for $\mathcal M(t)$:
	\begin{equation}\label{cak1}
		\mathcal M(t)=k\,t^{-\alpha}\;\;,
	\end{equation}
	where the parameters $k$ and $\alpha$ are the so-called \textit{line-force multiplier parameters} (or line-force parameters). \citet{abbott82} added a third line-force parameter called $\delta$, being the exponent of the diluted-electron number density, $N_e/W$ (where $W$ is the dilution factor). 
	With these three line-force parameters $(k,\alpha,\delta)$, $\mathcal M(t)$ becomes:
	\begin{equation}\label{M.C.ak}
		\mathcal M(t)=k\,t^{-\alpha}\left(10^{-11}\frac{N_e}{W}\right)^\delta\;\;.
	\end{equation}
	
	The physical interpretation of the line-force parameters \citep[see, e.g.][]{puls00} are as follows:
	\begin{itemize}
		\item The $k$ parameter is directly proportional to the effective number of driving lines, and is related to the fraction of the photospheric flux, which would have been blocked by all lines if they were optically thick and overlapping effects were not considered.
		Higher values of $k$ are obtained at higher densities and, therefore, higher mass-loss rates.
		In addition to the dependency of $\rho(r)$, $k$ presents also a strong dependence with metallicity and temperature due to the large number of driving lines: a lower temperature implies lower ionisation stages, and thus more lines; therefore, a higher $\mathcal M(t)$.
		More lines (above a given threshold line strength) are also present for higher metallicities.

		The overlapping of two or more spectral lines produces an overestimation in the calculated value of $k$. On the other hand, $k$ is underestimated when multiscattering effects are not taken into account (i.e., the summation in $\mathcal M(t)$ considers only direct photospheric radiation, and not radiation reprocessed in the wind).
		As was pointed out by \citet{puls87}, the inclusion of both effects might cancel, at least for O stars, and
		the \textit{effective} $k$ becomes moderately reduced. In this work, we have not considered these effects; therefore, our $k$ values should be maximum.
		
		\item The $\alpha$ parameter is related to the exponent of the line-strength distribution function, and quantifies also the ratio of the line acceleration from optically thick lines to the total one \citep[for details, see][]{puls08}.
		\item The $\delta$ parameter represents the change in the ionisation throughout the wind.
		It has been found that, high values of $\delta$ ($ \gtrsim 0.25$) "slows" the wind, yielding a different wind solution \citep{michel11}.
	\end{itemize}
	
	Some studies have pointed out that the line-force parameters are a function of radius \citep{schaerer94} or can be considered in a piecewise constant structure \citep{kudritzki02}. Nevertheless, in this work, we will consider $k$, $\alpha$ and $\delta$ as constants throughout the wind (see Section  \ref{constantakd}).

%_____METHODOLOGY_______________________________________________________________________________
	\section{Calculation of the $\mathcal M(t)$ factor \label{methods}}
	To calculate the $\mathcal M(t)$ factor, we include different improvements: (i) a larger line list, (ii) a quasi-NLTE approach for the ionisation equilibrium, (iii) a NLTE radiative stellar flux and (iv) an optical depth range in concordance with the wind structure.
	Then we test it for one single-step and also the whole iteration procedure until convergence of line-force parameters, velocity profile and mass-loss rate is achieved.
	
%_____Selection of atomic lines database________________________________________________________________	
	\subsection{Selection of atomic database}
		To calculate the line-acceleration and obtain a proper value of $\mathcal M(t)$, \citet{abbott82} established that it is necessary to sum the contribution of hundreds of thousands of spectral lines participating in the line-acceleration processes.
		Therefore, aiming to get the most accurate value of $\mathcal M(t)$, we decided to employ around $\sim 900\,000$ line transitions.
		These atomic data were obtained (and modified in format) from the atomic database list used by the code CMFGEN\footnote{Atomic data used here are those which were updated by DJH in 2016 (\url{http://kookaburra.phyast.pitt.edu/hillier/cmfgen_files/atomic_data_15nov16.tar.gz}).} \citep{hillier90,hillier98}.
		Specifically, we have extracted information related to energy levels, degeneracy levels, partition functions and oscillator strengths $f_l$, which are necessary to calculate the absorption coefficient $\eta_\text{line}$ of each line in terms of lower ($l$) and upper ($u$) level populations $n_l$ and $n_u$, and their statistical weights $g_l$ and $g_u$.
		The absorption coefficient $\eta_\text{line}$ is given by:
		\begin{equation}
			\eta_\text{line}=\frac{\pi e^2}{M.C.}f_l\frac{n_l}{\rho(r)}\left(1-\frac{n_u}{n_l}\frac{g_l}{g_u}\right)\;\;.
		\end{equation}
			
		Elements and ionisation stages considered in this work are listed in Table \ref{atomicdata}.
		\begin{table}
			\caption{\small{Atomic elements and ionisation stages used to calculate $\mathcal M(t)$.}}
			\label{atomicdata}
			\centering
			\begin{tabular}{rlrl}
				\hline\hline
				Elem. & Ions & Elem. & Ions\\
				\hline
				H & I & He & I$-$II\\
				Li & I$-$III & Be & I$-$IV\\
				B & I$-$V & C & I$-$IV\\
				N & I$-$VI & O & I$-$VI\\
				F & I$-$VI & Ne & I$-$VI\\
				Na & I$-$VI & Mg & I$-$VI\\
				Al & I$-$VI & Si & I$-$VI\\
				P & I$-$VI & S & I$-$VI\\
				Cl & I$-$VI & Ar & I$-$VI\\
				K & I$-$VI & Ca & I$-$VI\\
				Sc & I$-$VI & Ti & I$-$VI\\
				V & I$-$VI & Cr & I$-$VI\\
				Mn & I$-$VI & Fe & I$-$VI\\
				Co & I$-$VI & Ni & I$-$VI\\
				\hline
			\end{tabular}
		\end{table}
			
%_____Ionisation equilibrium____________________________________________________________________________
	\subsection{Ionisation equilibrium}
		Line-acceleration is calculated over the contribution of numerous transitions for every element at every ionisation stage present in the wind.
		\citet{abbott82} determined the ionisation degrees using the Saha's equation for extended atmospheres \citep{mihalas78}, namely:
		\begin{equation}\label{saha}
			\small{\left(\frac{N_{i+1}}{N_i}\right)_\text{LTE}=2\left(\frac{2\pi m_ek_B}{h^2}\right)^{3/2}\frac{T_R\sqrt{T_e}}{N_e/W}\frac{U_{i+1}}{U_i}e^{-\frac{E_i}{k_BT_e}}\;\;,}
		\end{equation}
		where $T_R$, $T_e$ are the radiation and electron temperatures, respectively, and $E_i$ is the ionisation energy from stage $i$ to $i+1$.
		More precise treatment called \textit{approximate NLTE} (hereafter quasi-NLTE) has been used by, e.g., \citet{mazzali93} and \citet{noebauer15}.
		Here the ionisation balance is determined by the application of the modified nebular approximation \citep{abbott85}.
		Following this treatment, the ratio of number densities for two consecutive ions can be expressed in term of its LTE value, multiplied by correction effects due to dilution of radiation field and recombinations:
		\begin{equation}\label{recombinated}
			\small{\frac{N_{i+1}}{N_i}\approx\left(\frac{N_e}{W}\right)^{-1}[\zeta_i+W(1-\zeta_i)]\sqrt{\frac{T_e}{T_R}}\left(\frac{N_{i+1}N_e}{N_i}\right)_\text{LTE}\,,}
		\end{equation}
		where $\zeta_i$ represents the fraction of recombination processes that go directly to the ground stage. Eq. (\ref{recombinated}) is an alternative description to the one given by \citet{puls05}, who included a different radiative temperature dependence in the wind, which is specially important in the far UV region of the spectrum that is not optically thick.
			
		Modifications in the treatment of atomic populations $X_i$, with $i$ being the excitation level, are also based on the work of \citet{abbott85}.
		It is necessary to make distinction between metastable levels (with no permitted electromagnetic dipole transitions to lower energy levels) and all the other ones:
		\[ \left(\frac{X_i}{X_1}\right) =
		\begin{cases}
		\left(\frac{X_i}{X_1}\right)_\text{LTE}       & \quad \text{metastable levels}\\
		\\
		W(r)\left(\frac{X_i}{X_1}\right)_\text{LTE}  & \quad \text{others}
		\end{cases}
		\]
		
		Atomic partition functions, $U_i$ (necessary for Saha's equation and the calculation of atomic populations), are calculated following the formulation of \citet{cardona10}, i.e.,:
		\begin{equation}\label{cardonapartition3}
			U_i=U_{i,0}+G_{jk}e^{-\varepsilon_{jk}/T}+\frac{m}{3}(n^3-343)e^{-\hat E_{n*jk}/T}\;\;,
		\end{equation}
		where $U_{i,0}$ are the constant partition functions, $\hat E_{n*jk}$ is the mean excitation energy of the last level of the ion, $n$ is the maximum excitation stage to be considered, while $G_{jk}$, $\varepsilon_{jk}$ and $m$ are parameters tabulated by \citet{cardona10}.
			
		The advantage of this treatment is that it provides values for atomic partition functions explicitly as a function of temperature and implicitly of electron density, giving a more accurate ionisation balance.
		Following \citet{noebauer15}, the temperature will be treated as a constant ($T_R=T_e=T_\text{eff}$).
		Then, for a specific value of $(T_\text{eff},N_e)$, the ratio between number densities of ionisation stages $j$ and  $i$ (for a specific $Z$-element) is calculated by a matrix (hereafter ionisation matrix) given by:
		\begin{equation}
			D_{Z,i,j}=\frac{N_j}{N_i}=\prod_{i\leq k< j}\frac{N_{k+1}}{N_k}\;\;.
		\end{equation}
			
		In reference to the abundances of the different chemical elements, these were adopted from the solar abundances given by \citet{asplund09}.
		However, these can be easily modified to evaluate stars with nonsolar metallicity (see Sect, \ref{results}).

		At this point, it is necessary to remark that previous authors  \citep{abbott82,noebauer15} have considered the diluted-electron density $N_e/W$ as constant throughout the wind.
		Nevertheless, to calculate $\delta$, $\mathcal M(t)$ must be evaluated considering changes in the ionisation stages, and therefore, $N_e(r)/W(r)$. Since, the calculation of electron density depends on  the ionisation stages of each specie which in turn are functions of $N_e$, we deal with a coupled nonlinear problem.
		To obtain a solution, we use the following formula to calculate (as an initial value) the electron number density:
		\begin{equation}
		\label{eqNe0}
			N_{e,0}=\frac{\rho(r)}{m_H}\,\frac{X_\text{H}+2X_\text{He}}{X_\text{H}+4X_\text{He}}\, ,
		\end{equation}
		with $m_H$ being the hydrogen atom mass, and $X_\text{H}$ and $X_\text{He}$ the abundances of hydrogen and helium, respectively.
			
		We used this initial electron density to start calculating the ionisation matrix and to recalculate both atomic populations and electron density iteratively:
		\begin{eqnarray}
			N_e(r)&=&\left(X_\text{H}\frac{D_{1,1,2}}{1+D_{1,1,2}}+X_\text{He}\frac{(D_{2,1,2}+2D_{2,1,3})}{1+D_{2,1,2}+D_{2,1,3}}\right)\nonumber\\
			& &\times \, \frac{\rho(r)}{X_\text{H}+4X_\text{He}}\,.
		\end{eqnarray}
			
		\begin{figure}
			\centering
			\includegraphics[width=\linewidth]{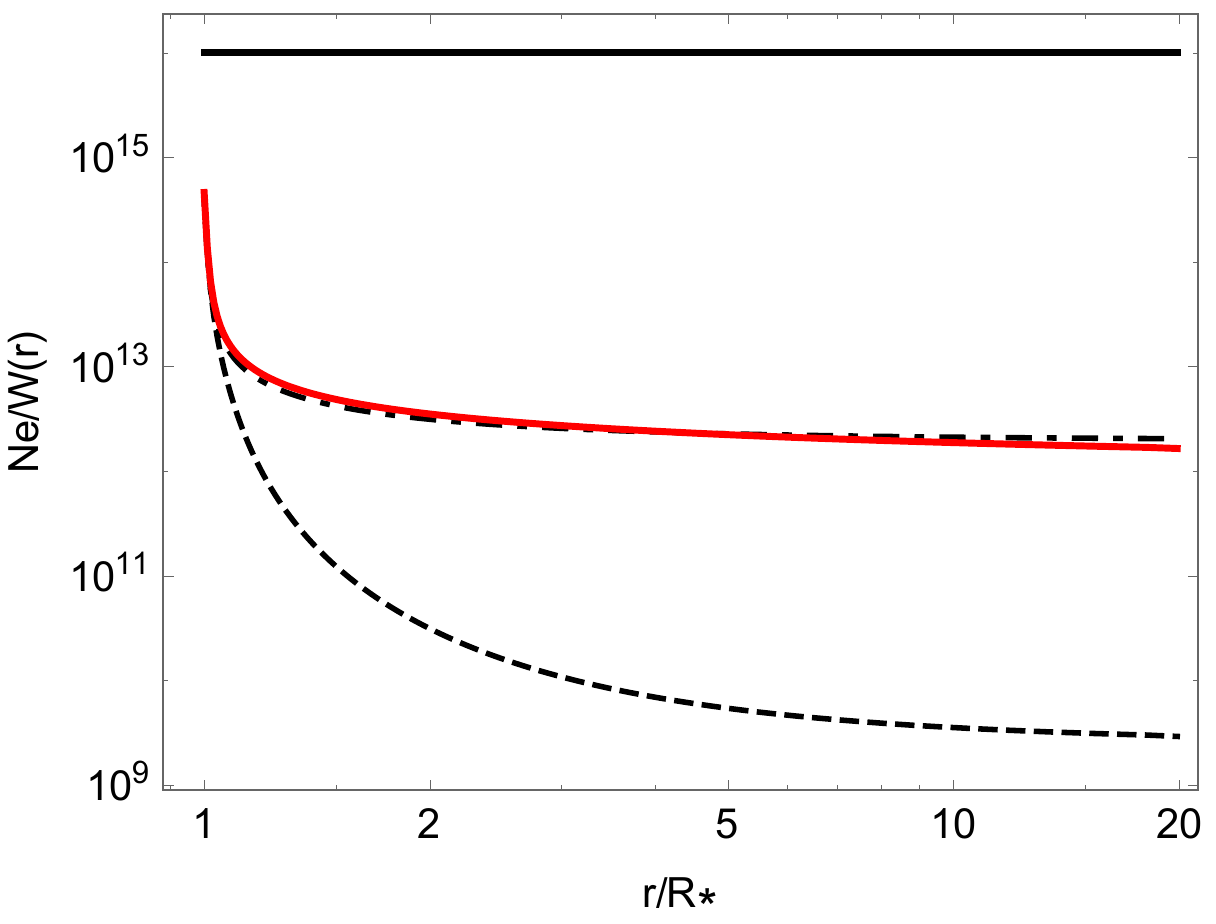}
			\caption{\small{Final value of $N_e/W(r)$ as function of stellar radius even when $N_{e,0}$ is set as constant input (black solid line), after one iteration (single dashed line), after two iterations (dashed-dotted line) and after five iterations (red solid line).}}
			\label{ne-conv}
		\end{figure}
		Convergence of $N_e$ is easily obtained after just a few iterations (see Fig. \ref{ne-conv}).
		It is important to remark that even when we use $N_{e,0}$ as a constant value (not described by Eq.~\ref{eqNe0}), the final converged value for $N_e$ is the same.

%_____Radiaton field__________________________________________________________________________________
	\subsection{Radiation field}
		Together with an accurate treatment of atomic populations and electron density, Eq. \ref{mt_definition} requires as an input the radiation field in the term $F_\nu/F$.
			
		\citet{abbott82} used the radiation fields from Kurucz' models \citep{kurucz79}, whereas \citet{noebauer15} from a blackbody. In this work, we use the radiation field calculated by the NLTE line-blanketing plane-parallel code \textsc{Tlusty} \citep{hubeny95,lanz03}.
		
		The overlap effects among tens of thousands of spectral lines are not considered when we sum the contributions to the force-multiplier $\mathcal M(t)$ across the wind.
		However, line blanketing effects are partially considered as we are using the \textsc{Tlusty} radiation field in the calculations of $\mathcal M(t)$.

%_____Determination of the optical depth range____________________________________________________________
	\begin{figure}[htbt]
		\centering
		\includegraphics[width=0.9\linewidth]{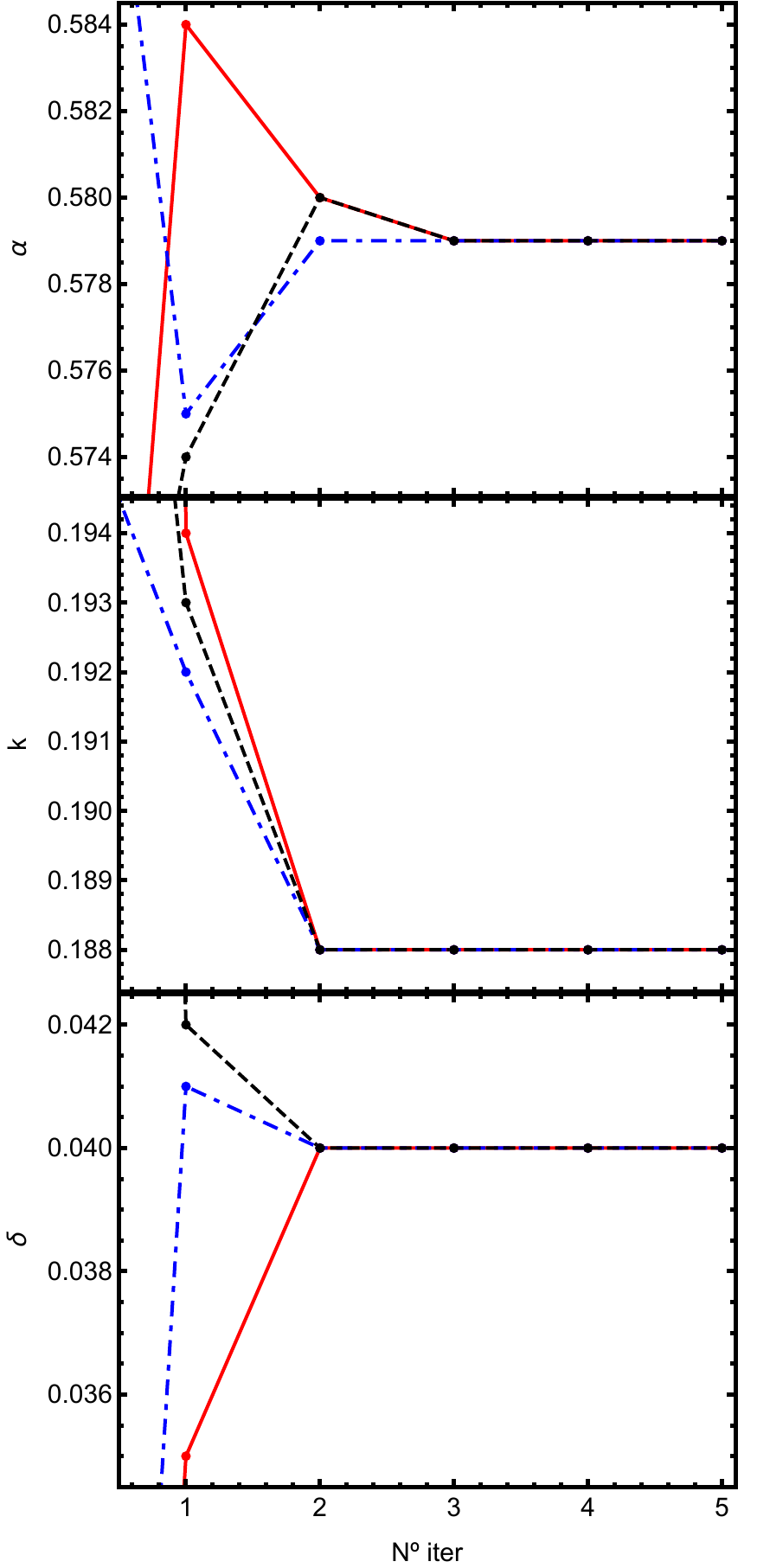}
		\caption{\small{Values of $\alpha$, $k$ and $\delta$ as a function of the iteration number, starting from different initial values. Different initial values (iteration 0, not shown) converge to the same final self-consistent solution.}}
		\label{iters}
	\end{figure}

	\begin{figure}[htbt]
		\centering
		\includegraphics[width=0.9\linewidth]{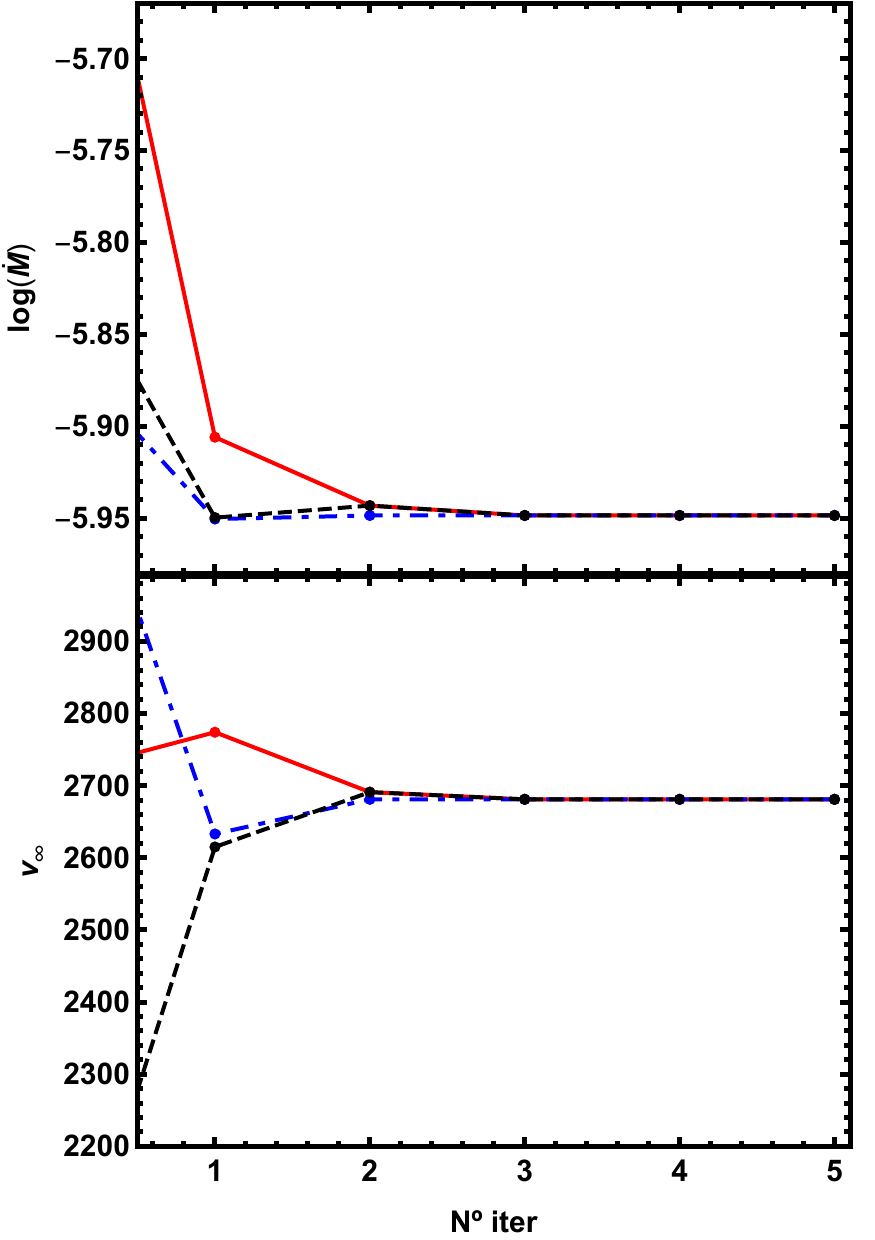}
		\caption{\small{Same as as Fig. \ref{iters}, but for the mass-loss rate and terminal velocity.}}
		\label{windpariters}
	\end{figure}

	\subsection{Determination of the optical depth range}
		Previous studies by \citet{abbott82} and \citet{noebauer15} have considered a fixed range for the optical depth $t$ to fit the force multiplier (Eq. \ref{M.C.ak}).
		
		However, given the definition of $t$ (Eq. \ref{t}), it is clear that the optical depth range is constrained by the physical properties of the stellar wind (density and velocity profiles).
		For this reason, calculations presented in this work are constrained inside the wind, characterised by this range of $t$.
		
		Because m-CAK theory is based upon Sobolev approximation \citep{sobolev60,lamersandcassinelli} in this work we will use as upper and lower limits of $t$ its values at the sonic point and at infinity (usually $r\sim 100\,R_*$), respectively.
		It is important to remark that although $t$ decreases outward it never reaches zero.
		Therefore, it is always possible to define a proper range.

%_____Iterative procedure______________________________________________________________________________
	\subsection{Iterative procedure}
		Velocity profile and $\dot M$ from hydrodynamics is required in order to calculate the line-acceleration $g_\text{line}$.
		At the same time, line-force parameters fitted from $g_\text{line}$, are necessary to solve the m-CAK hydrodynamic equations and obtain the mass-loss rate and velocity profile.
		Therefore, a self-consistent iterative procedure should be implemented to solve this coupled nonlinear problem.
		
		Our procedure is the following: (i) using a $\beta$-law profile with a given mass-loss rate, initial values for the line force parameters $(k_0,\alpha_0,\delta_0)$ are calculated; (ii) a numerical solution of the equation of motion (Eq. \ref{motion}) is obtained with  \textsc{HydWind \footnote{This code solves the m-CAK equation of motion with an eigenvalue that depends on the mass-loss rate. At the location of the singular point, both solution branches (singular point to stellar surface and singular point to infinity) are smoothly merged to obtain the velocity profile, see \citet{ppk,friend86} and \citet{michel04} for details.}}, getting an improved hydrodynamics: $v(r)$ and $\dot{M}$; (iii) a new force multiplier is calculated; (iv) new line-force parameters $(k_i,\alpha_i,\delta_i)$ are fitted from  $\mathcal M(t)$; and (v) steps ii - iv are iterated.
		Convergence is usually obtained after $\sim4-5$ iterations (see Fig. \ref{iters}), independently on the initial values.
		Our criterion for convergence is when  two consecutive iterations ($i$, $i+1$) get a value  for $\|\Delta p\|= \|p^{i+1}-p^{i}\| \,\leq 10^{-3}$, where $p$ is a line-force parameter and this condition should be satisfied for each one of these parameters.
		
		Figure \ref{windpariters} shows the convergence of the mass-loss rate (top panel) and the terminal velocity (lower panel) as a function of the procedure's iterations.
		Both values depend nonlinearly on the stellar and line-force parameters.
		
%_____One single iteration test_________________________________________________________________________	
	\begin{table}[htbp]
		\caption{\small{Comparison of $k$ and $\alpha$ parameters from Abbott (A) and Noebauer \& Sim (N), with our one single-step results.}}
		\label{initresults}
		\centering
		\begin{tabular}{ccclcccc}
			\hline
			\hline
			& & & & \multicolumn{2}{c}{Previous Studies} & \multicolumn{2}{c}{Present Work}\\
			\cline{5-6} \cline{7-8}
			& $T_\text{eff}$ & $N_e/W$ & $\delta$ & $k$ & $\alpha$ & $k_1$ & $\alpha_1$ \\
			& [kK] & [cm$^{-3}$] &\\
			\hline
			A & 30 & $1.0\times10^8$ & 0.12 & 0.093 & 0.576 & 0.062 & 0.661\\
			A & 30 & $1.0\times10^{11}$ & 0.12 & 0.156 & 0.609 & 0.097 & 0.611\\
			A & 30 & $1.0\times10^{14}$ & 0.12 & 0.571 & 0.545 & 0.487 & 0.450\\
			A & 40 & $1.8\times10^8$ & 0.12 & 0.051 & 0.684 & 0.072 & 0.639\\
			A & 40 & $1.8\times10^{11}$ & 0.12 & 0.174 & 0.606 & 0.120 & 0.609\\
			A & 40 & $1.8\times10^{14}$ & 0.12 & 0.533 & 0.571 & 0.289 & 0.552\\
			N & 42 & $1.0\times10^{15}$ & 0.0 & 0.381 & 0.595 & 0.376 & 0.572\\
			A & 50 & $3.1\times10^8$ & 0.092 & 0.089 & 0.640 & 0.148 & 0.611\\
			A & 50 & $3.1\times10^{11}$ & 0.092 & 0.178 & 0.606 & 0.196 & 0.595\\
			A & 50 & $3.1\times10^{14}$ & 0.092 & 0.472 & 0.582 & 0.289 & 0.566\\
			\hline
		\end{tabular}
	\end{table}

	\subsection{A single-step test}
		To compare our line-force parameters with the results obtained by \citet{abbott82} and \citet{noebauer15}, we use one single-step only.
		Following these authors, $\delta$ and $N_e/W$ are set as input and the optical depth range is fixed between $-6<\log t<-1$. The selection of a fixed interval of $\log t$ does not  require any velocity field structure.
		Furthermore, we have considered Kurucz' and  black-body fluxes to reproduce  \citet{abbott82} and \citet{noebauer15} calculations, respectively.
		Then, starting from a $\beta$-law and a $\dot{M}$, we calculate $k_1$ and $\alpha_1$ (single-step).
		These results are shown in Table \ref{initresults}.
		The  coefficients of determination, $R$-Squared, for $\alpha$ and $k$ (respectively) between previous and our single-iteration results are: (i) $R_{\alpha}^2=0.87$ and $R_k^2=0.93$ for $T_{\rm{eff}} \geq 40\,000$ K; (ii) $R_{\alpha}^2=0.4$ and $R_k^2=0.81$ for $T_{\rm{eff}} \geq 30\,000$ K.
		We conclude that our calculations reproduced previous results, now using a modern atomic database and abundances.
		
%_____LTE RESULTS__________________________________________________________________________________
\section{Results}\label{results}
	\begin{table*}[t]
		\centering
		\caption{\small{Self-consistent line-force parameters $(k,\alpha,\delta)$ for adopted standard stellar parameters, together with the resulting terminal velocities and mass-loss rates ($\dot M_\text{SC}$). Ratios between self-consistent mass-loss rates and Vink's recipe values \citep[re-scaled to match metallicity from][]{asplund09} using $ v_\infty/v_{\rm{esc}}=2.6$ are shown in the last column. Error margins for mass-loss rates and terminal velocities are derived over a variation of $\pm500$ for effective temperature, $\pm0.05$ for logarithm of surface gravity and $\pm0.1$ for stellar radius.}}
		\begin{tabular}{cccc|rc|ccc|cccc}
			\hline
			\hline
			$T_\text{eff}$ & $\log g$ & $R_*/R_\odot$ & $Z_/Z_\odot$ & $\log t_\text{in}$ & $\log t_\text{out}$ & $k$ & $\alpha$ & $\delta$ & $ v^\text{SC}_\infty$ & $\dot M_\text{SC}$\ & $\dot M_\text{SC}/\dot M_\text{Vink}$\\
			$[\text{kK}]$ & & & & & & & & & [km s$^{-1}$] & [$10^{-6}M_\odot\,\text{yr}^{-1}$]\\
			\hline
			45 & 4.0 & 12.0 & 1.0 & $-0.31$ & $-4.53$ & 0.167 & 0.600 & 0.021 & $3\,432\pm240$ & $2.0\pm_{0.5}^{0.65}$ & 1.00\\
			45 & 4.0 & 12.0 & 0.2 & $-0.77$ & $-4.85$ & 0.142 & 0.493 & 0.017 & $2\,329\pm210$ & $0.38\pm_{0.11}^{0.15}$ & 0.74\\
			45 & 3.8 & 16.0 & 1.0 & $0.28$ & $-4.07$ & 0.135 & 0.648 & 0.022 & $3\,250\pm300$ & $6.4\pm_{1.3}^{1.6}$ & 0.84\\
			45 & 3.8 & 16.0 & 0.2 & $-0.06$ & $-4.28$ & 0.114 & 0.545 & 0.014 & $2\,221\pm230$ & $1.7\pm_{0.45}^{0.6}$ & 0.88\\
			42 & 3.8 & 16.0 & 1.0 & $-0.10$ & $-4.36$ & 0.137 & 0.629 & 0.027 & $3\,235\pm300$ & $3.4\pm_{0.7}^{0.9}$ & 0.94\\
			42 & 3.8 & 16.0 & 0.2 & $-0.55$ & $-4.73$ & 0.108 & 0.534 & 0.019 & $2\,313\pm230$ & $0.73\pm_{0.21}^{0.3}$ & 0.79\\
			42 & 3.6 & 20.4 & 1.0 & 0.70 & $-3.80$ & 0.122 & 0.671 & 0.039 & $2\,738\pm230$ & $11\pm_{2.5}^{3.5}$ & 0.74\\
			42 & 3.6 & 20.4 & 0.2 & 0.37 & $-4.09$ & 0.091 & 0.586 & 0.022 & $2\,043\pm200$ & $3.1\pm_{0.75}^{1.2}$ & 0.82\\
			40 & 4.0 & 12.0 & 1.0 & $-0.88$ & $-4.97$ & 0.164 & 0.581 & 0.027 & $3\,300\pm220$ & $0.66\pm_{0.15}^{0.19}$ & 1.17\\
			40 & 4.0 & 12.0 & 0.2 & $-1.43$ & $-5.44$ & 0.133 & 0.492 & 0.038 & $2\,329\pm160$ & $0.11\pm_{0.03}^{0.05}$ & 0.76\\
			40 & 3.6 & 20.4 & 1.0 & $0.42$ & $-3.96$ & 0.118 & 0.659 & 0.044 & $2\,813\pm290$ & $6.6\pm_{1.4}^{1.8}$ & 0.89\\
			40 & 3.6 & 20.4 & 0.2 & $-0.05$ & $-4.40$ & 0.091 & 0.572 & 0.025 & $2\,116\pm230$ & $1.7\pm_{0.4}^{0.5}$ & 0.90\\
			40 & 3.4 & 18.0 & 1.0 & $1.30$ & $-3.14$ & 0.099 & 0.715 & 0.094 & $1\,548\pm240$ & $14.5\pm_{3.5}^{5.0}$ & 0.73\\
			40 & 3.4 & 18.0 & 0.2 & $1.90$ & $-3.50$ & 0.073 & 0.650 & 0.047 & $1\,334\pm230$ & $4.7\pm_{1.3}^{2.4}$ & 0.92\\
			38 & 3.8 & 16.0 & 1.0 & $-0.63$ & $-4.79$ & 0.130 & 0.610 & 0.036 & $3\,153\pm240$ & $1.2\pm_{0.25}^{0.3}$ & 1.10\\
			38 & 3.8 & 16.0 & 0.2 & $-1.18$ & $-5.28$ & 0.091 & 0.542 & 0.033 & $2\,473\pm300$ & $0.25\pm_{0.06}^{0.08}$ & 0.89\\
			36 & 4.0 & 12.0 & 1.0 & $-1.45$ & $-5.50$ & 0.132 & 0.580 & 0.036 & $3\,314\pm200$ & $0.21\pm_{0.05}^{0.065}$ & 1.17\\
			36 & 4.0 & 12.0 & 0.2 & $-1.97$ & $-5.97$ & 0.101 & 0.517 & 0.068 & $2\,402\pm140$ & $0.036\pm_{0.01}^{0.014}$ & 0.78\\
			36 & 3.6 & 20.4 & 1.0 & $-0.29$ & $-4.55$ & 0.104 & 0.644 & 0.062 & $2\,809\pm240$ & $2.2\pm_{0.5}^{0.7}$ & 1.12\\
			36 & 3.6 & 20.4 & 0.2 & $-0.87$ & $-5.09$ & 0.071 & 0.581 & 0.033 & $2\,534\pm220$ & $0.5\pm_{0.13}^{0.17}$ & 1.00\\
			36 & 3.4 & 18.0 & 1.0 & $1.78$ & $-3.77$ & 0.091 & 0.686 & 0.116 & $1\,708\pm170$ & $4.4\pm_{1.0}^{1.6}$ & 1.13\\
			36 & 3.4 & 18.0 & 0.2 & $0.41$ & $-4.21$ & 0.072 & 0.607 & 0.048 & $1\,566\pm160$ & $1.0\pm_{0.25}^{0.4}$ & 1.01\\
			34 & 3.8 & 16.0 & 1.0 & $-1.27$ & $-5.37$ & 0.103 & 0.604 & 0.043 & $3\,093\pm210$ & $0.34\pm_{0.07}^{0.1}$ & 1.12\\
			34 & 3.8 & 16.0 & 0.2 & $-1.93$ & $-5.94$ & 0.069 & 0.555 & 0.028 & $2\,791\pm180$ & $0.074\pm_{0.018}^{0.025}$ & 0.95\\
			34 & 3.6 & 20.4 & 1.0 & $-0.61$ & $-4.82$ & 0.095 & 0.637 & 0.074 & $2\,732\pm180$ & $1.2\pm_{0.3}^{0.4}$ & 1.25\\
			34 & 3.6 & 20.4 & 0.2 & $-1.29$ & $-5.46$ & 0.058 & 0.590 & 0.031 & $2\,642\pm180$ & $0.25\pm_{0.05}^{0.07}$ & 1.03\\
			32 & 3.4 & 18.0 & 1.0 & $0.37$ & $-4.30$ & 0.078 & 0.675 & 0.159 & $1\,653\pm190$ & $1.3\pm_{0.3}^{0.5}$ & 1.67\\
			32 & 3.4 & 18.0 & 0.2 & $-0.70$ & $-4.15$ & 0.053 & 0.610 & 0.052 & $1\,847\pm140$ & $0.23\pm_{0.05}^{0.075}$ & 1.16\\
			\hline
		\end{tabular}
		\label{standardtable}
	\end{table*}
	\begin{figure}[htbp]
		\centering
		\includegraphics[width=0.99\linewidth]{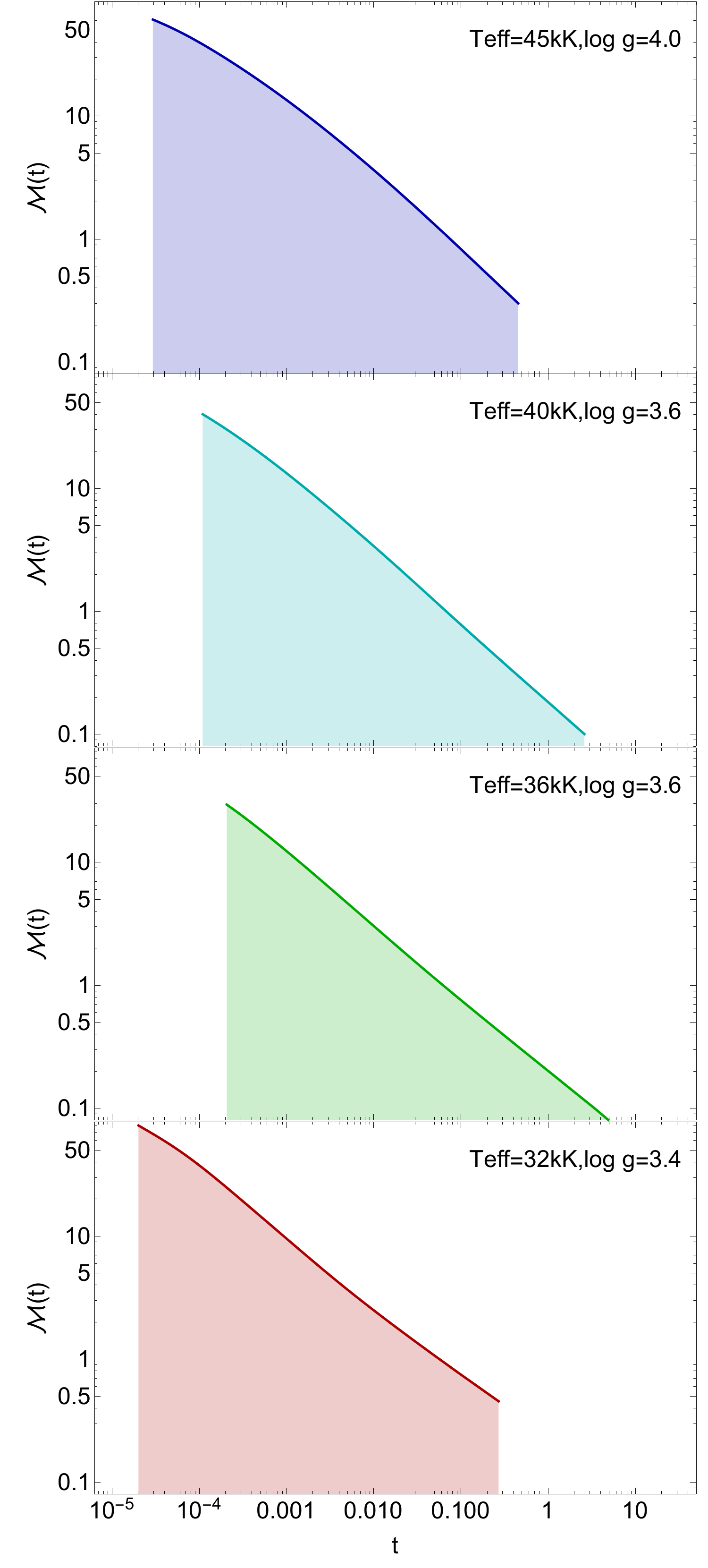}
		\caption{\small{Force-multiplier $\mathcal M(t)$ as function of $t$ for some stellar models presented on Table \ref{standardtable} with $T_\text{eff}=45\,000$ K and $\log g=4.0$ (blue), $T_\text{eff}=40\,000$ K and $\log g=3.6$ (cyan), $T_\text{eff}=36\,000$ K and $\log g=3.4$ (green) and $T_\text{eff}=32\,000$ K and $\log g=3.4$ (red). Coloured areas below curves indicate the range of $t$, where the fits for $(k,\alpha,\delta)$ have been adjusted.}}
		\label{fastplots}
	\end{figure}
	\begin{figure}[htbp]
		\centering
		\includegraphics[width=0.9\linewidth]{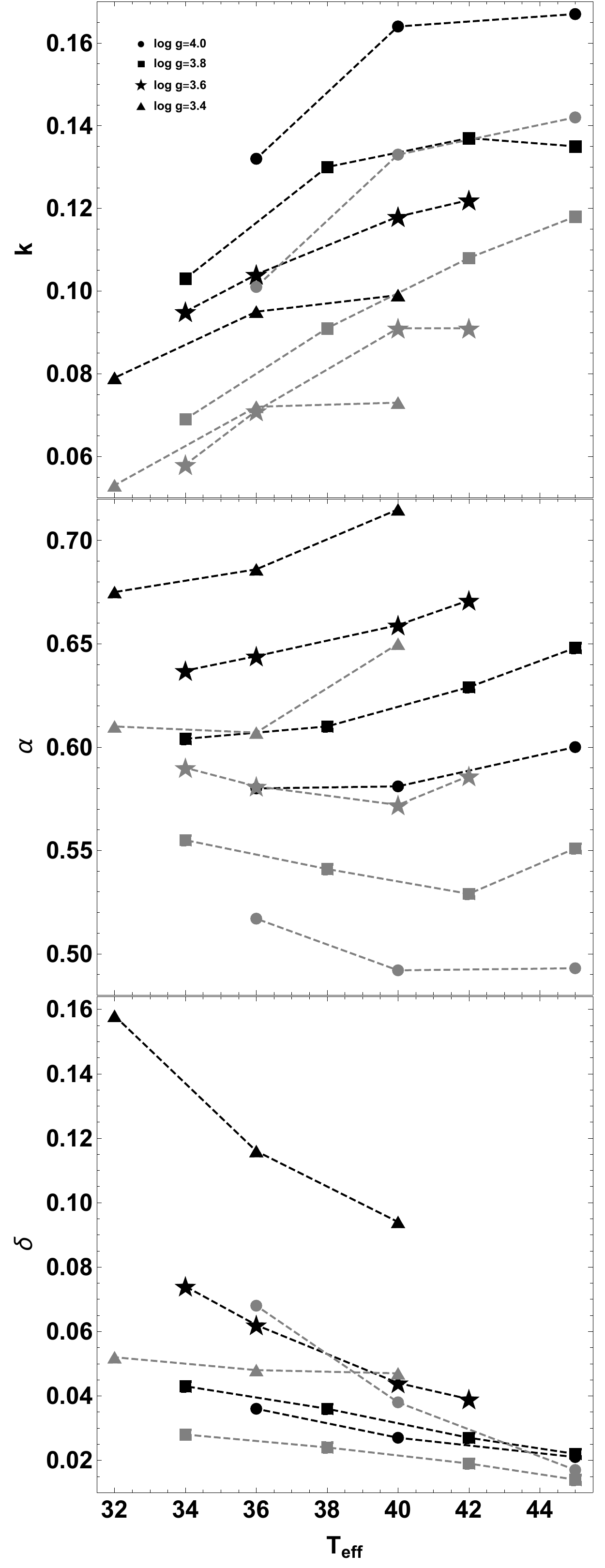}
		\caption{Behaviour of line-force parameters $(k,\alpha,\delta)$ as a function of the effective temperature (in kK), for different surface gravities and metallicities. Circles represent models with $\log\,g=4.0$, squares: $\log\,g=3.8$, stars: $\log\,g=~3.6$, and triangles:  $\log g=3.4$. Black dashed lines are for models with solar metallicity and grey dashed lines for $Z=Z_{\sun}/5$.}
		\label{behaviours}
	\end{figure}

%_____Self-consistent calculations_______________________________________________________________________
	\subsection{Self-consistent calculations}
		The following results are computed  self-consistently with the methodology detailed in Section \ref{methods}.

		Self-consistent solutions for a grid of models are presented in Table \ref{standardtable}.
		The effective temperature ranges from 32 to 45 kK and $\log\, g$ from $3.4$ to $4.0$ dex.
		This grid considers different stellar radii and two abundances: 1 and 1/5 of the solar value.
		This table shows the stellar parameters, the calculated $t$-range, and the fitted m-CAK line-force.
		In addition, we calculated the corresponding wind solution using \textsc{HydWind}, and their error margins were derived considering variations of $\Delta T_{\rm{eff}}=\pm500$, $\Delta\log g=\pm0.05$, and $\Delta R_*=\pm0.1R_\odot$ in the stellar radius, keeping constant the line-force parameters.
		
		Convergence has been checked for each solution. Figure \ref{fastplots} shows  the last iteration of $\mathcal M(t)$ for four models from Table \ref{standardtable} with different ranges of $t$.
		Due to the quasi-linear behaviour of the logarithm of the force-multiplier, parameters $k$ and $\alpha$ are easily fitted and their values can be considered constant throughout the wind (see Sect. \ref{constantakd}).
		To fit $\delta$ in the $\mathcal M(t)$--$N_e/W$ plane, it is necessary to perform an extra calculation of $\mathcal M(t)$ using a slightly different value for the diluted-electron density.
		Last column of this table shows the ratio between our mass-loss rate and the one calculated using Vink's recipe \citep{vink01}, with $v_{\infty}/v_{esc}=2.6$ and rescaled to current abundances \citep{asplund09}.
		The mean value of $\dot M_\text{SC}/\dot M_\text{Vink} = 0.98 \pm 0.2$.
		As we have not included in our procedure multi-line nor line-overlapping processes, we support conclusion given by \citet{puls87} that these effects are somewhat canceled, because we do not observe relevant discrepancies in the mass-loss rates when a comparison with Vink's recipe is performed.
		
		In Fig. \ref{behaviours}, we observe clear trends for the behaviour of the $(k,\alpha,\delta)$ parameters with $T_{\rm{eff}}$,  $\log\,g$, and $Z$.
		While $k$ increases and $\delta$ decreases as a function of the effective temperature, for both metallicities.
		It is interesting to remark the influence of the surface gravity on the resulting line-force parameters, values for $k$ and $\delta$ decrease as the gravity decreases.
		Notice that globally our line-force parameter results are similar to the values obtained in previous works \citep{puls00,kudritzki02,noebauer15}. However, we found an important dependence on $\log g$ as a result of the hydrodynamic coupling in the self-consistent procedure.
		
		On the other hand, the behaviour of $\alpha$ depends on the metallicity, it increases with effective temperature for solar abundance, but for low abundance and low gravities, it slowly decreases with temperature.
		Moreover, the change in $\alpha$ is more significant for $\log g$ than for $T_{\rm{eff}}$: a difference in $\Delta \log g \pm 0.2$ dex produces a $\Delta \alpha \sim0.04$, whereas variations on $\Delta T_\text{eff} =\pm\, 2\,000\,$K, might produce $\Delta \alpha \sim 0.02$.
				
\begin{figure}[htbp]
			\centering
			\includegraphics[width=0.9\linewidth]{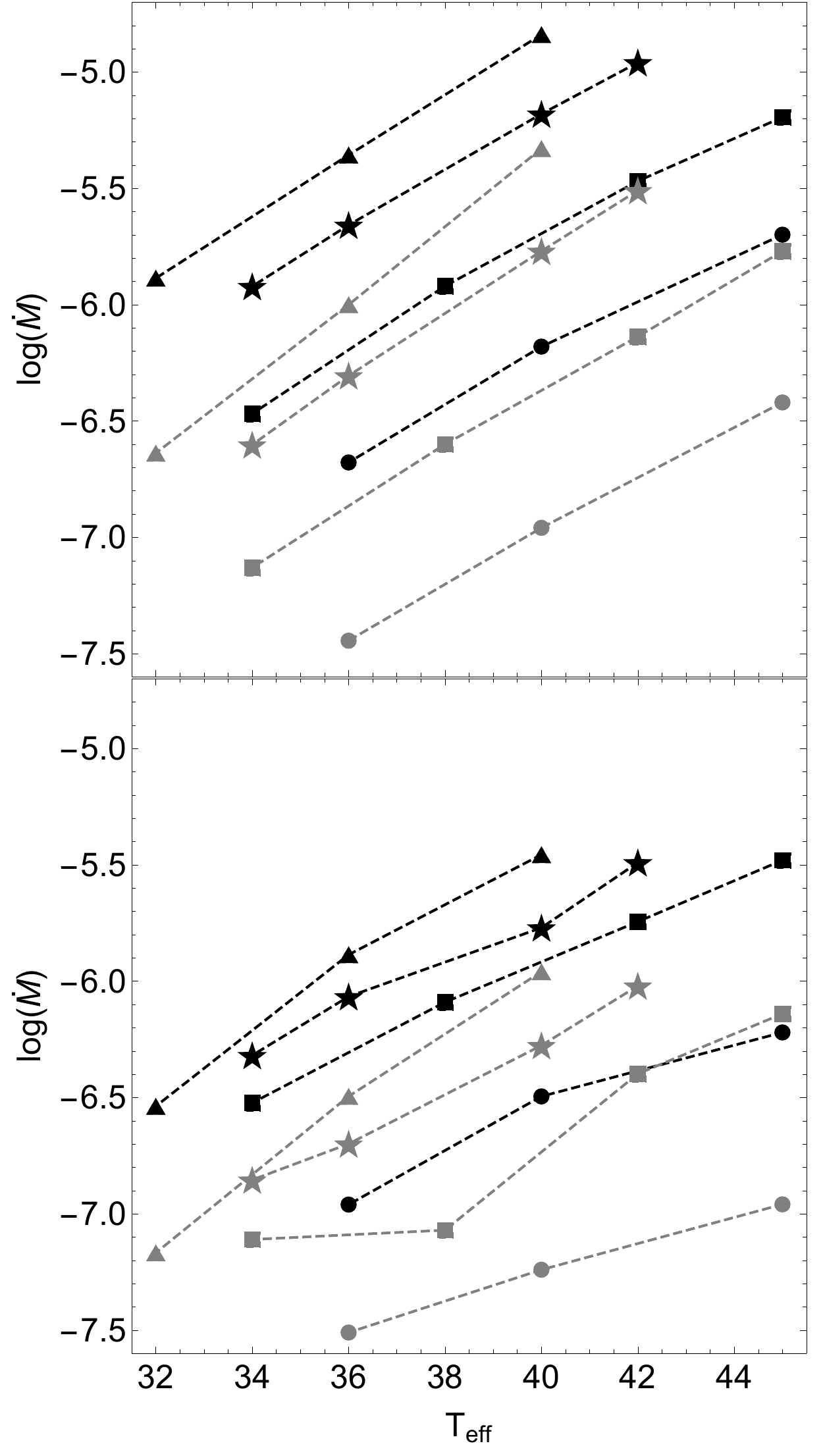}
			\caption{Behaviour of mass-loss rate as a function of effective temperature (in kK) for different abundances and gravities. Top panel is for self-consistent calculations and bottom panel is for Abbott's procedure, now including the finite-disk correction factor. Symbol description is the same as than in Fig.~\ref{behaviours}.}
			\label{windbehaviours}
		\end{figure}

		Figure \ref{windbehaviours} shows the results for the mass-loss rates as a function of the effective temperature, for different gravities and metallicities.
		The upper panel shows the results from our self-consistent procedure and the bottom panel shows the result using Abbott's methodology (a single iteration) to calculate line-force parameters and apply them in our hydrodynamic code \textsc{HydWind} (hereafter Abbott's procedure).
		We found that $\dot{M}$ increases with effective temperature and metallicity and decreases with gravity.
		This behaviour is similar to the one obtained using Abbott's procedure, but the self-consistent calculated mass-loss rates are about $30\%$ larger. 
		
		From the mass-loss results tabulated in Table \ref{standardtable}, a simple relationship for solar-like metallicities (with a  coefficient of determination or $R$-squared, $R^2=0.999$) reads:
		\begin{align}
			\log \dot M_{Z=1.0}=&10.443\times\log\left(\frac{T_\text{eff}}{1000\text{ K}}\right)\nonumber\\
			&-1.96\times\log g\nonumber\\
			&+0.0314\times(R_*/R_\odot)\nonumber\\
			&-15.49\nonumber,\\
		\end{align}
		and for metallicity $Z/Z_\odot=0.2$ the relationship reads (with $R^2=0.999$):
		\begin{align}
			\log \dot M_{Z=0.2}=&11.668\times\log\left(\frac{T_\text{eff}}{1000\text{ K}}\right)\nonumber\\
			&-2.126\times\log g\nonumber\\
			&+0.04\times(R_*/R_\odot)\nonumber\\
			&-17.63\nonumber,\\
		\end{align}
		where $\dot M$ is given in $10^{-6}M_\odot\,\text{yr}^{-1}$.
		These relationships could be considered analogous to that given by \citet{vink00} to obtain theoretical mass-loss rates for solar-like metallicities. However, the advantage of our description is that it depends only on \textit{stellar parameters} and we do not need to consider the value of $v_{\infty}/v_{esc}$. 
		It is important to remark, however, that this formula has been derived for the following ranges:
		\begin{itemize}
			\item $T_\text{eff}=32-45$ kK
			\item $\log g=3.4-4.25$
			\item $M_*/M_\odot\ge25.0$.
		\end{itemize}
		
		\begin{figure}[htbp]
			\centering
			\includegraphics[width=0.9\linewidth]{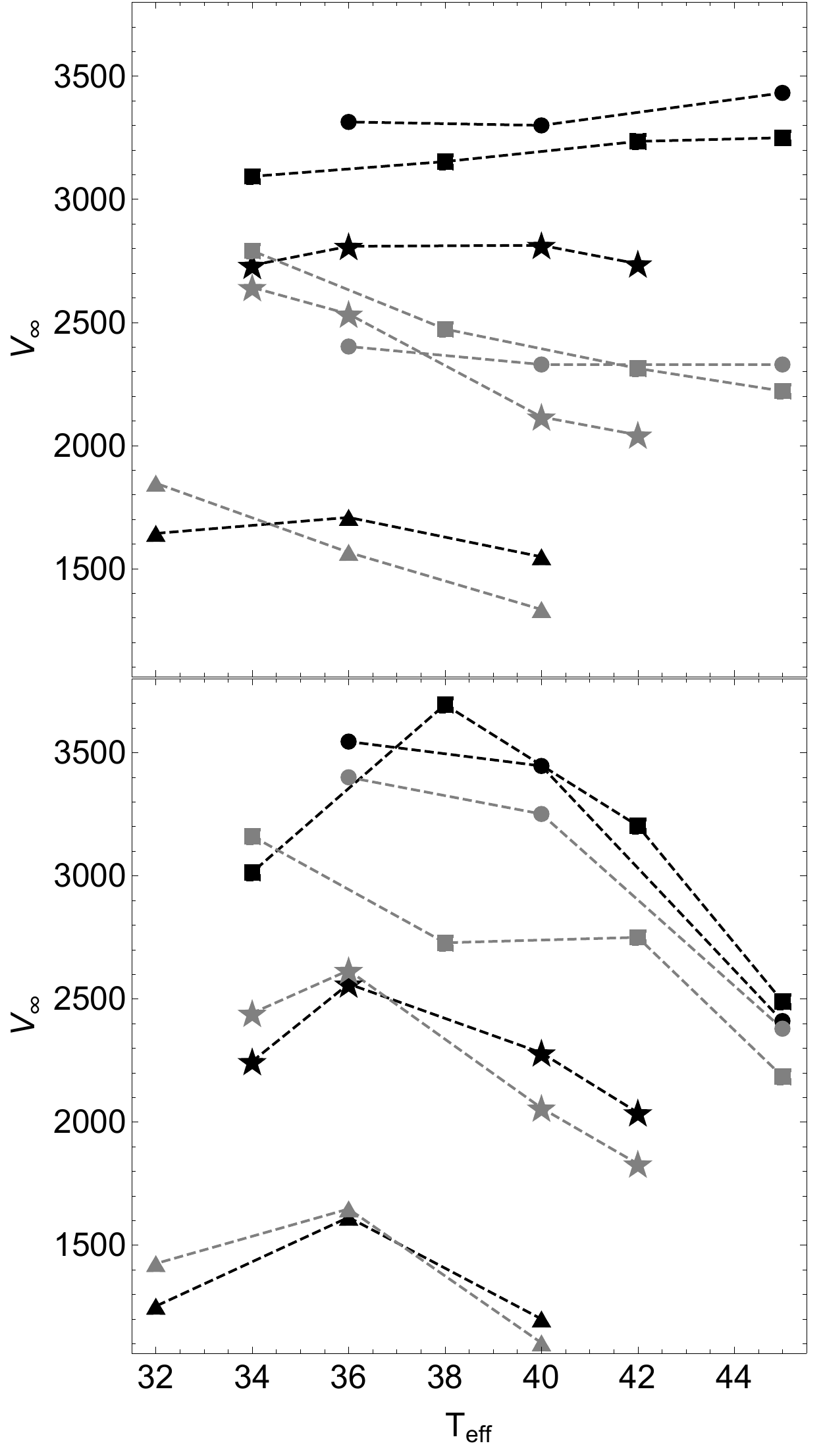}
			\caption{Same as Fig. \ref{windbehaviours}, but for the terminal velocities.}
			\label{wind2behaviours}
		\end{figure}
					
		Concerning terminal velocities, see Fig. \ref{wind2behaviours}, self-consistent calculations (top panel) show that $v_\infty$ is almost constant with respect to the effective temperature, but it decreases as a function of $\log g$ and $Z$.
		On the other hand, Abbott's procedure results do not show the same behavior and exhibit a maximum in the $T_{\rm{eff}}$ interval.

%Analysis of errors___________________________________________________________________________________
	\subsection{Range of validity for line-force parameters}\label{constantakd}
		\begin{table*}[htbp]
			\centering
			\caption{Influence of the optical depth interval on the line-force parameters for some reference models given in Table \ref{standardtable}. Absolute values of the differences in the resulting wind parameters with respect to the reference solution are presented.}
			\begin{tabular}{ccrc|ccc|rr}
				\hline\hline
				$T_\text{eff}$ & $\log g$ & $\log t_\text{in}$ & $\log t_\text{out}$ & $k$ & $\alpha$ & $\delta$ & $|\Delta v_\infty|$ & $|\Delta\dot M|$\\
				& & & & & & & [km s$^{-1}$] & [$10^{-6}M_\odot\,\text{yr}^{-1}$]\\
				\hline
				45\,000 & 4.0 & $-0.31$ & $-2.03$ & 0.099 & 0.686 & 0.037 & $780$ & $0.23$\\
				& & $-0.31$ & $-2.87$ & 0.107 & 0.650 & 0.029 & $600$ & $0.30$\\
				& & $-0.31$ & $-3.71$ & 0.120 & 0.638 & 0.027 & $420$ & $0.21$\\
				& & $-0.31$ & $-4.55$ & 0.167 & 0.600 & 0.021 & $0$ & $0$\\
				\hline
				40\,000 & 4.0 & $-0.87$ & $-2.50$ & 0.099 & 0.633 & 0.040 & $521$ & $0.09$\\
				& & $-0.87$ & $-3.32$ & 0.099 & 0.634 & 0.036 & $610$ & $0.07$\\
				& & $-0.87$ & $-4.14$ & 0.107 & 0.621 & 0.026 & $594$ & $0.07$\\
				& & $-0.87$ & $-4.96$ & 0.164 & 0.581 & 0.027 & $0$ & $0$\\
				\hline
				40\,000 & 3.6 & 0.08 & $-1.44$ & 0.095 & 0.666 & 0.090 & $247$ & $0.58$\\
				& & 0.08 & $-2.28$ & 0.098 & 0.680 & 0.075 & $75$ & $0.13$\\
				& & 0.08 & $-3.12$ & 0.101 & 0.692 & 0.067 & $323$ & $0.92$\\
				& & 0.08 & $-3.96$ & 0.118 & 0.659 & 0.044 & $0$ & $0$\\
				\hline
				36\,000 & 3.6 & $-0.29$ & $-2.00$ & 0.084 & 0.637 & 0.112 & $520$ & $0.58$\\
				& & $-0.29$ & $-2.85$ & 0.092 & 0.648 & 0.078 & $114$ & $0.15$\\
				& & $-0.29$ & $-3.70$ & 0.089 & 0.668 & 0.075 & $267$ & $0.01$\\
				& & $-0.29$ & $-4.55$ & 0.104 & 0.644 & 0.062 & $0$ & $0$\\
				\hline
				32\,000 & 3.4 & $0.37$ & $-1.49$ & 0.066 & 0.630 & 0.251 & $631$ & $0.77$\\
				& & $0.37$ & $-2.43$ & 0.075 & 0.636 & 0.221 & $457$ & $0.57$\\
				& & $0.37$ & $-3.37$ & 0.079 & 0.662 & 0.179 & $168$ & $0.11$\\
				& & $0.37$ & $-4.31$ & 0.078 & 0.675 & 0.159 & $0$ & $0$\\
				\hline
			\end{tabular}
			\label{variationofcak}
		\end{table*}

		It is important to remember that the range of optical depths used to calculate our self-consistent  line-force parameters is defined along almost all the atmosphere of the star, i.e., downstream from the sonic point.
		This procedure improves  the criterion used by \citet{abbott82}, who determined the parameters at $t=10^{-4}$. 		
		This value sometimes lays outside the optical depth range here defined, as shown in Fig. \ref{fastplots}.

		To analyse the change on the line-force parameters due to the selection of the $t$-range, we define four different intervals inside the whole range of $t$, and compute these parameters in each range. Table \ref{variationofcak} summarises these calculations.
		 Regarding the uncertainties of our procedure in the terminal velocities, these are of the same order as the uncertainties owe to the errors in the determination of the stellar parameters in the range $32\,000$ K $ < T_{\rm{eff}} < 40\,000$ K, while, the uncertainties in $\dot{M}$ are much lower than the ones produced by variations of stellar parameters. These small uncertainties indicate that it is a good approximation to consider line-force parameters as constants throughout the wind.
		Due to the fact that the entire $t$-range represents the physical conditions of almost all the wind, we recommend using the complete optical depth range to derive the line-force parameters.
		
		For $T_\text{eff} < 30\,000$ K, we found that $\log\,\mathcal M(t)$ is no longer linear with respect to $\log t$ and the corresponding line-force parameters can be approximated to a linear piecewise description.
		Due to this reason, we establish that our set of self-consistent solutions describes stellar winds for effective temperatures and $\log g$ in the range $32\,000-45\,000$ K and $3.4-4.0$ dex, respectively.

% Calculated spectra
	\section{Synthetic spectra}\label{syntheticspectra}
			
		In order to know whether our calculations reproduce realistic physical features observed in hot stars, we calculate  synthetic spectra for three O-type stars using FASTWIND.
		We select some stars in the range of the considered $T_{\rm{eff}}$, trying to cover the extreme cases of temperature and $\log g$.
		We choose firstly the O4 I(n)fp star $\zeta$-Puppis (HD 66811) because it has been extensively studied \citep{puls96,repolust04,puls06,sota11,bouret12,noebauer15}.
		Mentioned authors have independently adopted different sets of stellar and wind parameters, which are summarised in Table \ref{zpuppispar}.
		Here, the wind parameters were determined by \citet{repolust04}. \citet{puls06} has used their parameters and derived clumped mass-loss rates from H$\alpha$, IR and radio, using  analytical expressions for the corresponding opacities, whereas \citet{bouret12} used CMFGEN.
		Both calculations include clumping, so these results correspond to a clumped mass-loss rate.\footnote{FASTWIND uses the clumping factor $f_\text{cl}\ge 1$ (with $f_\text{cl} = 1$ representing the smooth limit), where $f_\text{cl} = 1/f$ if the inter-clump medium was void \citep[][]{sundqvist18}. On the other hand, CMFGEN-clumping is represented by the so-called volume filling factor $f$, which scales homogeneous and clumped mass-loss rates under the relationship $\dot M_\text{hom}=\dot M_\text{clump}/\sqrt{f}$ (notice that this $f$ takes values between 0 and 1).}
		On the other hand, the mass-loss rate given by \citet{noebauer15} was obtained using their Monte-Carlo radiation hydrodynamics (M.C.RH) method assuming a homogeneous media ($f_\text{cl}=1.0$).
			
		Particularly, we compare our results with those given by \citet{puls06}, who did an exhaustive analysis of the clumping throughout the wind.
		Two different values for  mass-loss rate are given by these authors, because they considered different stellar radii depending on the assumed distance for $\zeta$-Puppis: (i) the "conventional" ($d=460$ pc) and (ii) the one given by \citet[][$d=730$ pc]{sahu93}.
		We examine here the "conventional" case with $R_*/R_\odot=18.6$.
		We can observe from Table \ref{zpuppispar} (last row), that our new calculated  mass-loss rate agree quite well with the value from \citet{puls06}.

		\begin{table*}[htpb]
			\centering
			\caption{\small{Stellar and wind parameters for $\zeta$-Puppis from previous studies compared with our  self-consistent results. Line-force parameters are also listed.}}
			\begin{tabular}{lccccc|ccccc}
				\hline
				\hline
				\multicolumn{6}{c}{Previous Studies} & \multicolumn{5}{c}{Present Work}\\
				Reference & $T_\text{eff}$ & $\log g$ & $R_*/R_\odot$ & $\dot M$ & $v_\infty$ & $k$ & $\alpha$ & $\delta$ & $\dot M_\text{SC}$ & $v^\text{SC}_{\infty}$\\
				& $[\text{kK}]$ & & & [$10^{-6}M_\odot\,\text{yr}^{-1}$] & [km\,s$^{-1}$] & & & & [$10^{-6}M_\odot\,\text{yr}^{-1}$] & [km\,s$^{-1}$]\\
				\hline
				\small{\citet{noebauer15}} & 42 & 3.6 & 19.0 & $45.0$ & 881 & 0.120 & 0.678 & 0.041 & $11.0\pm_{3.0}^{3.5}$ & $2\,500\pm280$\\
				\small{\citet{bouret12}} & 40 & 3.64 & 18.7 & $2.0$ & $2\,300$ & 0.120 & 0.655 & 0.039 & $5.2\pm_{1.2}^{1.6}$ & $2\,700\pm300$\\
				\multirow{ 2}{*}{\small{\citet{puls06}}} & 39 & 3.6 & 29.7 & $8.5$ & $2\,250$ & 0.115 & 0.654 & 0.044 & $9.3\pm_{2.2}^{2.9}$ & $3\,200\pm350$\\
				& 39 & 3.6 & 18.6 & $4.2$ & $2\,250$ & 0.114 & 0.658 & 0.049 & $4.6\pm_{1.1}^{1.3}$ & $2\,570\pm300$\\
				\hline
			\end{tabular}
			\label{zpuppispar}
		\end{table*}
			
		Figure \ref{comp413540} shows the observed spectra (kindly provided by D. J. Hillier) and the resulting synthetic spectra for $\zeta$-Puppis.
		Stellar parameters are taken from \citet[][see Table \ref{zpuppispar}]{puls06} and wind parameters from our self-consistent procedure ($\dot M_\text{SC}=4.6\times10^{-6}$ $M_\odot\,\text{yr}^{-1}$).
		We calculated three synthetic spectra with different clumping factors: $f_\text{cl}=1.0$ (homogeneous), $f_\text{cl}=5.0$ and $f_\text{cl}=9.0$.
		The best fit is for $f_\text{cl}=5.0$, which is the same clumping factor found by \citet{puls06} with their $\dot M=4.2\times10^{-6}$ $M_\odot$ yr$^{-1}$.
		Moreover, we also include the synthetic spectra obtained with the self-consistent solution (see Fig. \ref{comp000809}), calculated using the stellar parameters given by \citet[][see Table \ref{zpuppispar}]{bouret12} and \citet{najarro11}.
		The best fit is achieved when we use a clumping factor of $f_\text{cl}=5.0$.
		These results suggest that the real stellar parameters lie in the neighbourhood given by \citet{puls06} and \citet{najarro11}.
		
		The observed spectrum for  HD 163758 (O9 I) has been obtained from the UVES-POP database\footnote{\url{http://www.eso.org/sci/observing/tools/uvespop/field_stars_uptonow.html}}.
		We calculated the synthetic spectra for this star (see Fig. \ref{comp537366}) using stellar parameters from \citet{bouret12} and wind self-consistent parameters (see Table \ref{hd163758par}) with different clumping factors, the best fit is for $f_\text{cl}=6.0$. 
		\begin{table*}[htpb]
			\centering
			\caption{\small{Same as Table \ref{zpuppispar}, but for  
			HD 163758 and HD 164794. Stellar and wind parameters are  from \citet{bouret12} and \citet{krticka15} respectively.}}
			\begin{tabular}{lccccc|ccccc}
				\hline
				\hline
				\multicolumn{6}{c}{Previous Studies} & \multicolumn{5}{c}{Present Work}\\
				Name & $T_\text{eff}$ & $\log g$ & $R_*/R_\odot$ & $\dot M$ & $v_\infty$ & $k$ & $\alpha$ & $\delta$ & $\dot M_\text{SC}$ & $v^\text{SC}_{\infty}$\\
				& $[\text{kK}]$ & & & [$10^{-6}M_\odot\,\text{yr}^{-1}$] & [km\,s$^{-1}$] & & & & [$10^{-6}M_\odot\,\text{yr}^{-1}$] & [km\,s$^{-1}$]\\
				\hline
				HD 163758 & 34.5 & 3.41 & 21.0 & 1.6 & 2\,100 & 0.087 & 0.679 & 0.112 & $3.3\pm_{0.8}^{1.1}$ & $2\,040\pm280$\\
				HD 164794 & 43.8 & 3.92 & 13.1 & 2.9 & 3\,090 & 0.141 & 0.614 & 0.020 & $2.3\pm_{0.5}^{0.6}$ & $3\,304\pm400$\\
				\hline
			\end{tabular}
			\label{hd163758par}
		\end{table*}
		
		Last spectrum corresponds to the O3.5 V star HD 164794, also obtained from the UVES-POP database.
		Stellar parameters were extracted from \citet{krticka15}, as shown in Table \ref{hd163758par}.
		Contrary to previous cases, the best fit is obtained for the homogeneous model ($f_\text{cl}=1.0$, see Fig. \ref{comp858687}).

		In view of these first results, our self-consistent iterative procedure takes us quickly into the neighborhood of the solution that reproduces the observed wind spectra for O-type stars.

%_____DISCUSSION__________________________________________________________________________________
\section{Discussion}\label{discussion}
	We have developed a self-consistent methodology to calculate the line-force parameters and derived consequently mass-loss rates and velocity profiles.
	We found that mass-loss rate is about $30\%$ larger than the one obtained using Abbott's procedure (non-self-consistent calculation).

%Terminal velocity________________________________________________________________________________________
	\subsection{Terminal velocity}
		It is well known that the scaling relation for the terminal velocity in the frame of the line-driven wind theory. This  relation \citep{puls08} reads:
		\begin{equation}
			\label{cinfinit}
			v_{\infty} \approx 2.25 \, \sqrt{\frac{\alpha}{1-\alpha}} \, v_{\rm{esc}}\,.
		\end{equation}
		This is an approximation of the formula found by \citet[][their Eqs. 62 to 65]{kppa89}.

\begin{figure}[htbp]
			\centering
			\includegraphics[width=0.95\linewidth]{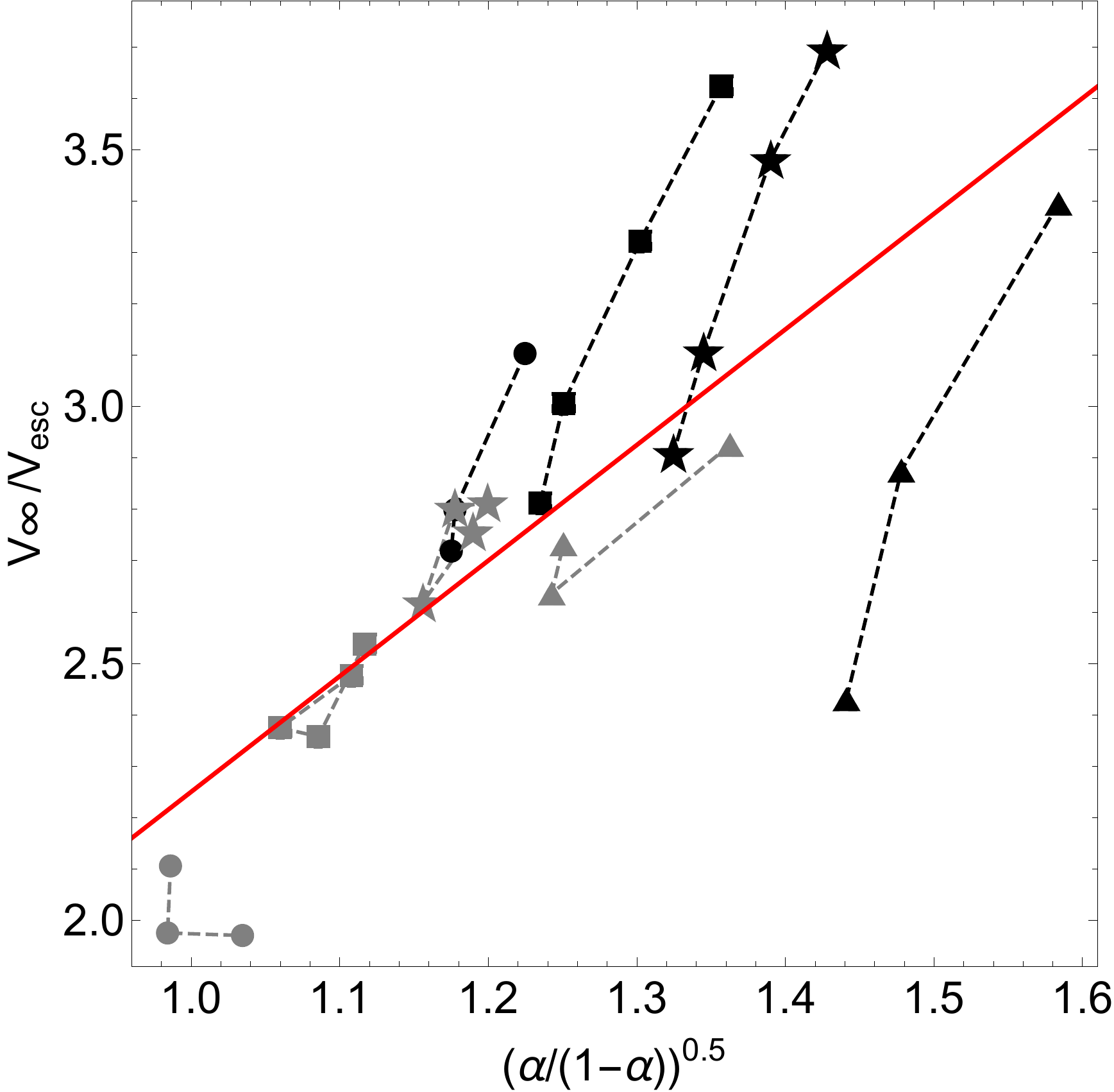}
			\caption{$v_{\infty}/v_{\rm{esc}}$ versus $\sqrt{\alpha/(1-\alpha)}$. For each set of $\log\,g$ values there is a linear dependence for $Z_\sun$. Slope 2.25 of Eq. \ref{cinfinit} is also displayed. For sub-solar abundance there is a unique linear relationship (see text for details). Symbol description is the same as in Fig. \ref{behaviours}}
			\label{vinfvesc}
		\end{figure}

		In Fig. \ref{vinfvesc} we have plotted $v_{\infty}/v_{\rm{esc}}$ versus $\sqrt{\alpha/(1-\alpha)}$ using the results from Table \ref{standardtable}.
		Contrary to the expected result (Eq. \ref{cinfinit}) for solar abundances, we find a different linear behaviour that strongly depends on the value of $\log\,g$. This is a new result that comes from applying our self-consistent procedure. The m-CAK equation of motion shows an interplay between the gravity ($\log\,g$) and the line force term. This balance of forces defines the location of the singular point and therefore fixes the value of $\dot{M}$. As a consequence, the velocity profile depends also on the value of $\log\,g$. This result cannot be obtained from Eq.\ref{cinfinit} which is an oversimplification of this nonlinear coupling. However, Eq. \ref{cinfinit} presents a fair fit when $Z$=$Z_{\sun}/5$, where the dependence of the slope on $\log g$ is weak because the radiation force is driven by fewer ions.

		The dependence of $v_{\infty}/v_{\rm{esc}}$ on $\log\,g$ yield that stars with  solar abundances present an intrinsic variations of $v_{\infty}/v_{\rm{esc}}$  in the range of $2.4 - 3.7$, as shown in Fig. \ref{vinfvesc}.
		This range might explain the scatter observed on the hot side of the bistability jump shown by \citet[][in their Fig. 12]{markova08}.

%Mass-loss rate______________________________________________________________________________________
	\subsection{Mass-loss rate}
		In this section we want to compare our theoretical results with the ones obtained from line-profile fittings for homogeneous (unclumped) winds with a $\beta$-law, and the mass-loss (recipe) from \citet{vink00}.
		\begin{table*}[htpb]
			\centering
			\caption{\small{Comparison of self-consistent with $\beta$-law (single-step) models for the two stars analyzed by \citet{bouret05}. Self-consistent models reproduce better the line-fitted wind parameters obtained by these authors ($\beta$=$1$: $v_{\infty}=3000\,$ km s$^{-1}$, $\dot{M}=1.8\times10^{-6}\,$ $M_\odot\,\text{yr}^{-1}$, and $\beta$=$0.8$: $v_{\infty}=2300\,$ km s$^{-1}$, $\dot{M}=6\times10^{-6}\,$ $M_\odot\,\text{yr}^{-1}$).}}
			\begin{tabular}{lccc|ccc|cc}
				\hline
				\hline
				Model&$T_\text{eff}$ & $\log g$ & $R_*/R_\odot$ & $k$ & $\alpha$ & $\delta$ & $v_\infty$ & $\dot M$\\
				& [kK] & & & & & & [km s$^{-1}$] & [$10^{-6}M_\odot\,\text{yr}^{-1}$]\\
				\hline
				Self-Consistent& 43.5 & 4.0 & 11.9 & 0.159 & 0.603  & 0.032 & $3\,342\pm240$ & $1.55\pm_{0.3}^{0.45}$ \\
				$\beta=1.0$ & 43.5 & 4.0 & 11.9  & 0.118 & 0.647 & 0.021 & $4\,187\pm290$ & $1.45\pm_{0.25}^{0.35}$ \\
				\hline
				Self-Consistent & 39 & 3.6 & 19.45 & 0.116 & 0.657 & 0.079 & $2\,412\pm210$ & $5.8\pm_{1.3}^{2.0}$ \\
				$\beta=0.8$ & 39 & 3.6 & 19.45  & 0.039 & 0.815 & 0.062 & $6\,789\pm570$ & $4.2\pm_{0.7}^{0.9}$ \\
				\hline
			\end{tabular}
			\label{tablebouret05atlas}
		\end{table*}

		Table \ref{tablebouret05atlas} shows our results for the only two O-type star reported by \citet{bouret05}: HD 96715, $T_\text{eff}=43.5$ kK, $\log g=4.0$, and  HD 1904290A, $T_\text{eff}=39$ kK, $\log g=3.6$.
		These results were obtained for the self-consistent solution together with the ones after just one iteration starting from a $\beta$-law.
		It is observed that models starting from a $\beta$-law largely overestimate the terminal velocity and slightly underestimate the mass-loss rate.
		Self-consistent calculations find a fairly good agreement to both: the observed mass-loss rate and terminal velocity.
		For the mass-loss rate in this figure, we have included the result calculated using \citet{vink00} recipe.
		It is clear that our self-consistent method gives values of $\dot{M}$ much closer to the observed ones.
		\begin{table*}[htpb]
			\centering
			\newcolumntype{d}{D{,}{,}{-1} }
			\caption{\small{Resulting self-consistent wind parameters ($v^\text{SC}_\infty$ and $\dot M_\text{SC}$) calculated for stars analyzed by \citet{markova18}. Error margins presented here for wind parameters are undergone from uncertainties of $\pm1\,000$ for $T_\text{eff}$ and $\pm0.1$ for $\log g$. Last two columns show the ratio between self-consistent and observed mass-loss rates and the ratio between self-consistent and  Vink's  mass-loss rates.}}
			\begin{tabular}{lllc|ccc|cc|cc}
				\hline
				\hline
				Field Star & $T_\text{eff}$ & $\log g$ & $R_*/R_\odot$ & $k$ & $\alpha$ & $\delta$ & $v^{SC}_\infty$ & \multicolumn{1}{c}{$\dot M$} & $\dot M_\text{SC}/\dot M_\text{obs}$ & $\dot M_\text{SC}/\dot M_\text{Vink}$\\
				& [kK] & & & & & & [km s$^{-1}$] & \multicolumn{1}{c}{$[10^{-6}M_\odot\,\text{yr}^{-1}]$} & &\\
				\hline
				HD 169582 & 37 & 3.5 & 27.2 & 0.102 & 0.668 & 0.063 & $3\,017\pm700$ & $7.1\pm_{2.4}^{3.6}$ & 1.10 & 1.26\\
				CD-43 4690 & 37 & 3.61 & 14.1 & 0.105 & 0.653 & 0.058 & $2\,310\pm540$ & $1.5\pm_{0.55}^{0.9}$ & 1.22 & 1.16\\
				HD 97848 & 36.5 & 3.9 & 8.2 & 0.123 & 0.601 & 0.034 & $2\,532\pm470$ & $0.17\pm_{0.06}^{0.09}$ & 0.89 & 0.95\\
				HD 69464 & 36 & 3.51 & 20.0 & 0.099 & 0.664 & 0.076 & $2\,412\pm580$ & $3.2\pm_{1.2}^{1.9}$ & 1.14 & 1.30\\
				HD 302505 & 34 & 3.6 & 14.1 & 0.092 & 0.643 & 0.077 & $2\,331\pm460$ & $0.68\pm_{0.26}^{0.42}$ & 1.24 & 0.98\\
				\hline
				HD 148546 & 31 & 3.22 & 24.4 & 0.073 & 0.718 & 0.243 & $1\,300\pm350$ & $5.3\pm_{2.5}^{4.7}$ & 0.94 & 2.24\\
				HD 76968a & 31 & 3.25 & 21.3 & 0.071 & 0.711 & 0.248 & $1\,212\pm300$ & $3.5\pm_{1.7}^{3.3}$ & 1.43 & 2.11\\
				HD 69106 & 30 & 3.55 & 14.2 & 0.068 & 0.644 & 0.149 & $1\,455\pm300$ & $0.21\pm_{0.09}^{0.16}$ & 1.48 & 1.78\\
				\hline
			\end{tabular}
			\label{tablemarkova18}
		\end{table*}
		\begin{figure}[htbp]
			\centering
			\includegraphics[width=0.9\linewidth]{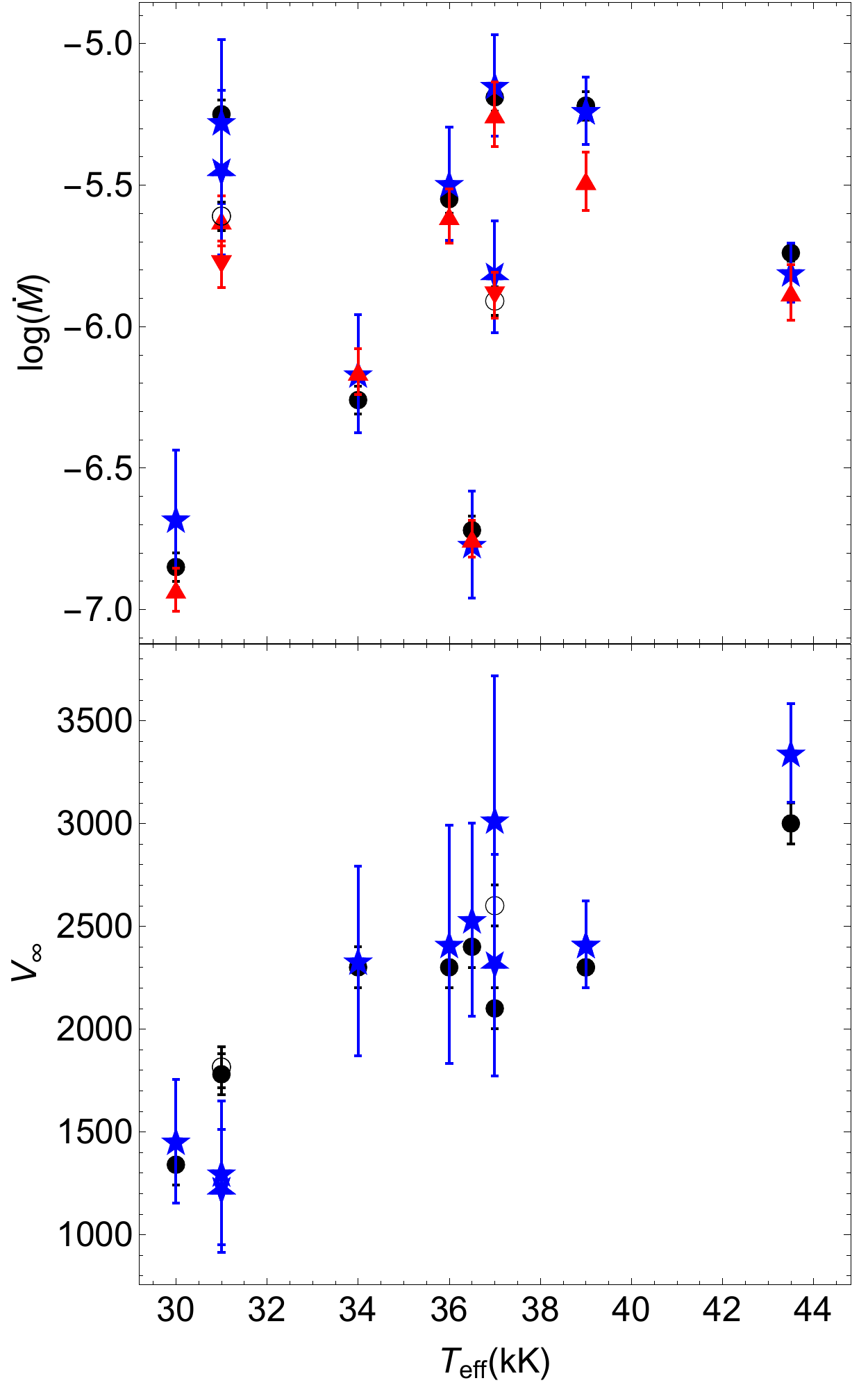}
			\caption{Comparison of mass-loss rates (upper panel) and terminal velocities (lower panel) as a function of the effective temperature. Blue stars correspond to results from this work, black disks to \citet{bouret05} and \citet{markova18} results, and red triangles to theoretical values from \citet{vink00}. The same colour code but with modified symbols (inverted blue stars, unfilled black circles and inverted red triangles) are used to represent Markova's stars with the same effective temperature but higher surface gravity.}
			\label{masslossbouret05}
		\end{figure}
			
		We also apply our self-consistent procedure to objects analyzed by means of FASTWIND adopting unclumped winds.
		For that purpose, we also examine some field Galactic O-type stars from \citet{markova18}.
		Table \ref{tablemarkova18} summarizes our results.
		We found a fair agreement between observed and calculated mass-loss rates (see Fig. \ref{masslossbouret05}).
		These results confirm that our methodology delivers the proper mass-loss rate for the ranges in  $T_\text{eff}$ and $\log g$ given above.
		Below these thresholds, mass-loss rates present larger values compared with both: observational and  Vink's theoretical values.
		This is probably due to the fact that the line-force multiplier is not longer a linear function of $t$ (in the $\log$-$\log$ plane, see Fig. \ref{fastplots}), and the line-force parameters are not constant throughout the wind.
						
		However, it is important to remark that uncertainties of $\Delta T_{\rm{eff}} \sim \pm1\,000$ K and $\Delta \log g \sim \pm0.1$ dex, produce uncertainties in the mass-loss rates up to a factor of 2 (see blue error bar in the top panel of Fig. \ref{masslossbouret05}), which can be considered as the upper threshold for the mass-loss rate.
		Hence, even though our self-consistent hydrodynamics gives confident values for $\dot M$, these good results are strongly dependent on the assumed stellar parameters.

%_____CONCLUSIONS________________________________________________________________________________
\section{Conclusions}\label{conclusions}
	In the present work we have presented a treatment to calculate a self-consistent line-force parameters coupled with the hydrodynamics in the frame of the radiation driven wind theory.
	Thanks to this procedure, we achieve a unique well-converged solution that does not depend on the chosen initial values.
	This is important because it reduces the number of free parameters (now $\beta$, $v_\infty$ and $\dot{M}$ are no more input parameters) to be determined by fitting synthetic spectra against observed ones.

	Our calculations contemplate the contribution to the line-force multiplier from more than $\sim 900\,000$ atomic transitions, an NLTE radiation flux from the photosphere and a quasi-LTE approximation for the occupational numbers.
	We have to notice that for $T_{\rm{eff}} > 30\,000$ K the line force parameters can be confidently used as constants throughout the wind. 

	The set of solutions given in Table \ref{standardtable} differs from previous line-force parameter calculations performed by \citet{abbott82} and \citet{noebauer15}.
	With these new values, we found a different scale relation for the terminal velocity that is steeper than the usually accepted one.
	This new relation might explain the observed scatter found in the terminal velocity from massive stars located at the hot side of the bistability jump \citep{markova08}.

	Concerning the wind parameters derived from modelling O-type stars with homogeneous winds, our mass-loss rates are in better agreement with the predicted ones given by the \citet{vink00} formula. 
	
	For the calculation of synthetic spectra for O-type stars ($\zeta$-Puppis, HD 163758 and HD 164794), we conclude that our procedure's values for mass-loss rate and hydrodynamics reproduce the observed line profiles when an adequate value for the clumping factor is chosen.
	
	Even knowing the limitations of the m-CAK theory, this remains an extremely useful framework to get an approach about the real parameters of stellar winds on massive stars.
	In spite of the approximations assumed under this theory, we obtain  reliable values for mass-loss rates and self-consistent hydrodynamics in a short period of time with a great CPU time savings (compare with big efforts made by, e.g., \citealt{mokiem05} or \citealt{fierro18}).

	Our new self-consistent procedure can be used to derive accurate mass-loss rates and: (i) study evolutionary tracks, where a high precision on terminal velocities is not required, and (ii) derive trusty clumping factors via line-profile fittings.
	
\begin{acknowledgements}
	We sincerely thank J. Puls for helpful discussions that improved this work and  for having put to our disposal his code FASTWIND. We thank the anonymous referee for his/her useful comments. We are very grateful to  D.~J. Hillier for allowing us to use CMFGEN-atomic-data and providing us with the observed spectrum of $\zeta$-Puppis.
	A.C.G.M. has been financially supported by the PhD Scholarship folio Nº 2116 1426 from National Commission for Scientific and Technological Research of Chile (CONICYT).
	A.C.G.M. is also thankful for support from the Chilean Astronomical Society (SOCHIAS).
	A.C.G.M. and M.C. acknowledge support from Centro de Astrofísica de Valparaíso.
	M.C. and L.S.C. are thankful for support from the project CONICYT+PAI/Atracción de Capital Humano Avanzado del Extranjero (Folio PAI80160057).
	L.S.C. acknowledges financial support from the Universidad Nacional de La Plata (Programa de Incentivos G11/137), the CONICET (PIP 0177), and the Agencia Nacional de Promoción Científica y Tecnológica (Préstamo BID, PICT 2016/1971), Argentina.
	R.O.J.V. is thankful for financial support from the UNLP under program PPID/G004.
	This project has received funding from the European Union’s Framework Programme for Research and Innovation Horizon 2020 (2014-2020) under the Marie Sk\l{}odowska-Curie grant Agreement No. 823734.
\end{acknowledgements}	

\software{\textsc{HydWind} \citep{michel04}, CMFGEN \citep{hillier90,hillier98,hillier01}, \textsc{Tlusty} \citep{hubeny95}, FASTWIND \citep{santolaya97,puls05}}

%_____BIBLIOGRAFÍA_______________________________________________________________________________
\bibliography{apj-paper_v0128} % your references Yourfile.bib
\bibliographystyle{aasjournal} % style aa.bst
%\end{multicols}

%_____APÉNDICE__________________________________________________________________________________
\appendix
\section{FASTWIND spectra}\label{append}
		\begin{figure*}[htbp]
			\centering
			\includegraphics[width=\linewidth]{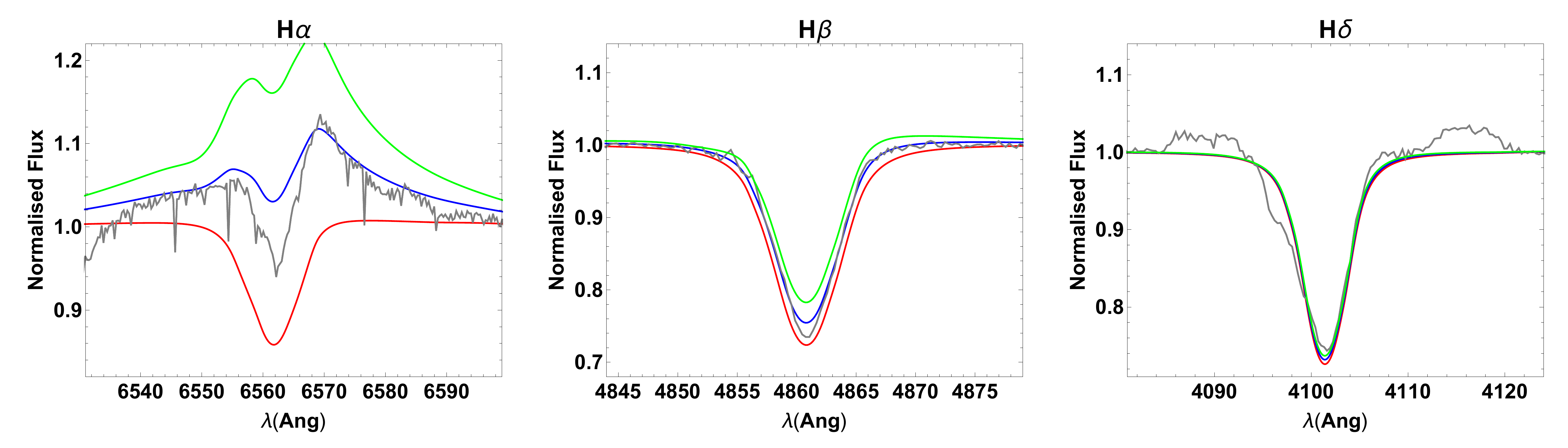}
			\includegraphics[width=\linewidth]{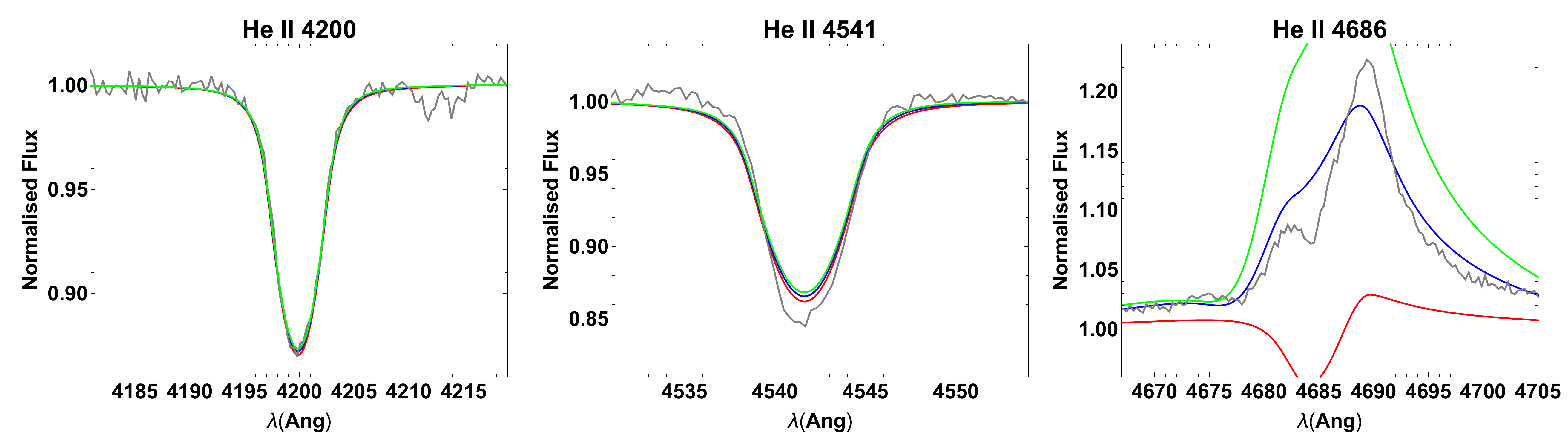}
			\includegraphics[width=\linewidth]{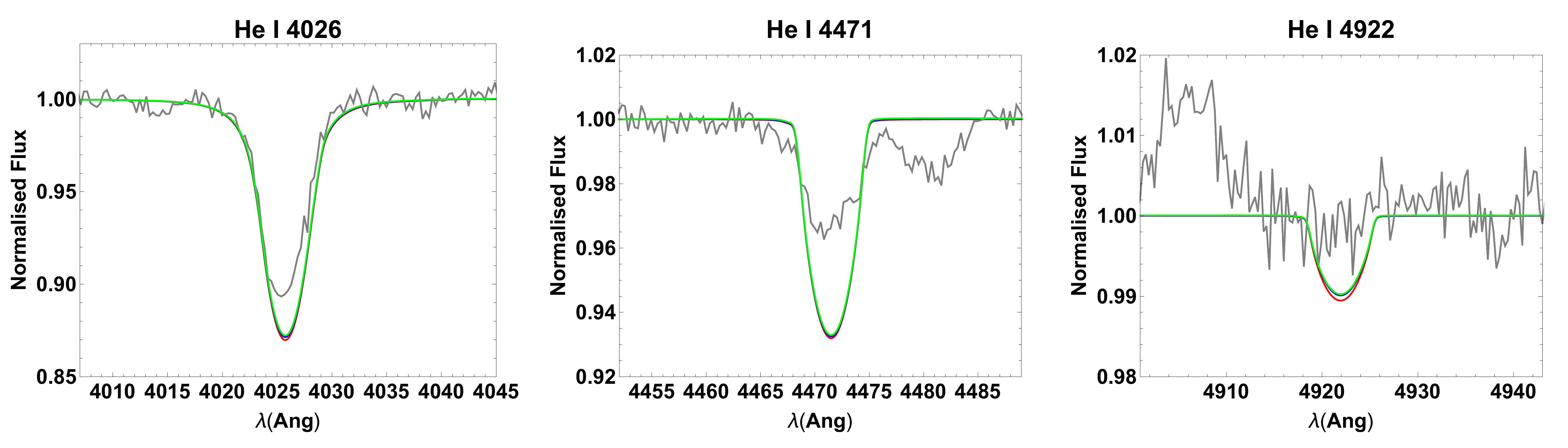}
			\caption{Resulting FASTWIND spectra for $\zeta$-Puppis with $T_\text{eff}=39$ kK, $\log g=3.6$, $R_*/R_\odot=18.6$ and $\dot M=4.6\times10^{-6}$ $M_\odot$ yr$^{-1}$. Clumping factors are $f_\text{cl}=1.0$ (red, homogeneous), $f_\text{cl}=5.0$ (blue) and $f_\text{cl}=9.0$ (green).}
			\label{comp413540}
		\end{figure*}
		\begin{figure*}[htbp]
			\centering
			\includegraphics[width=\linewidth]{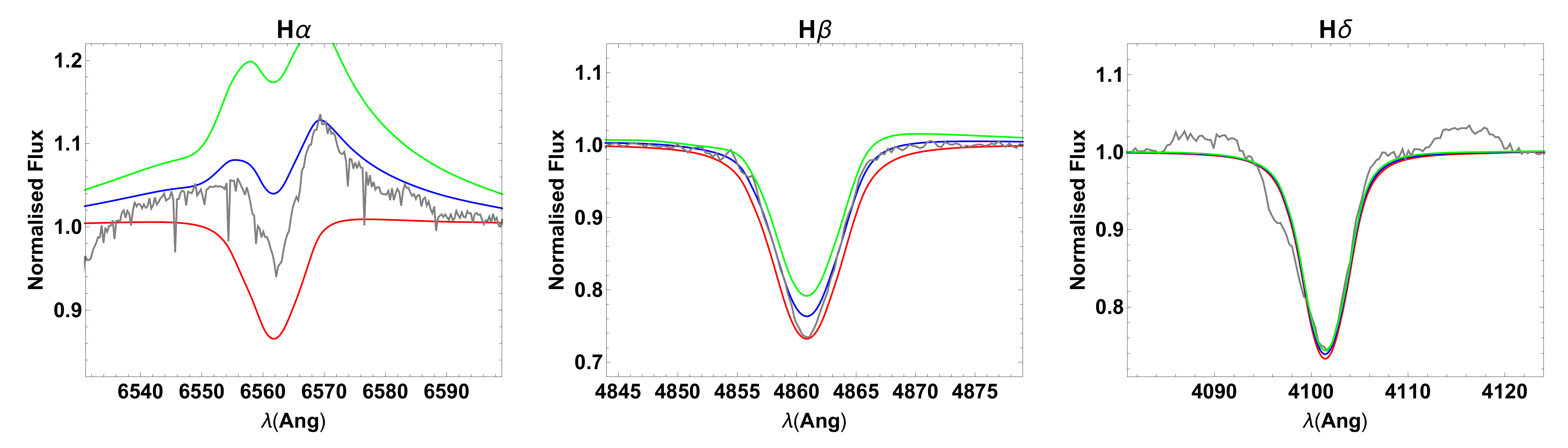}
			\includegraphics[width=\linewidth]{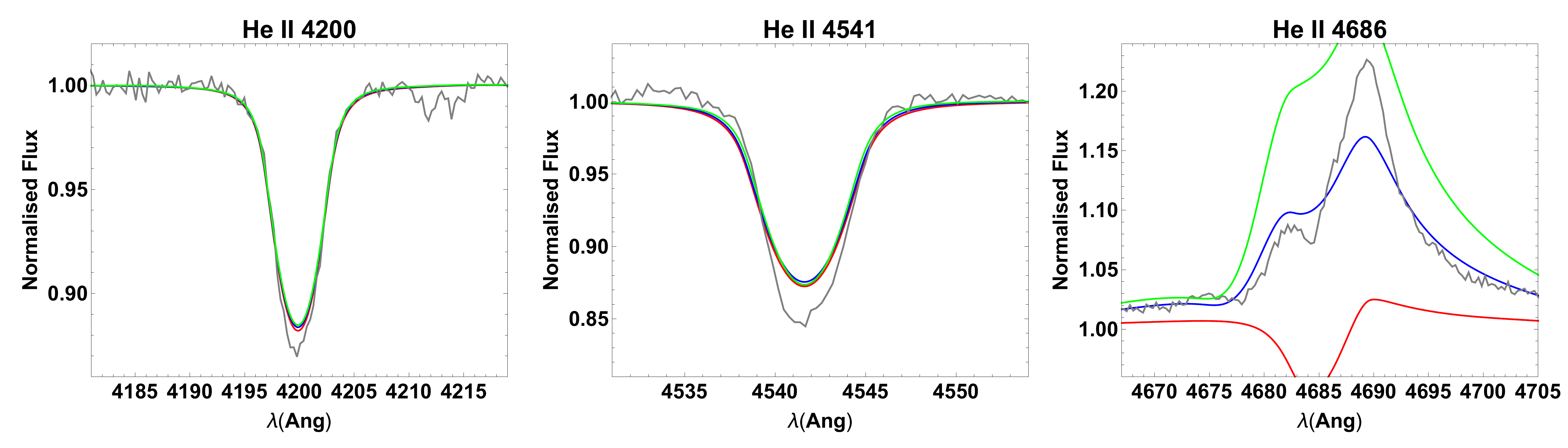}
			\includegraphics[width=\linewidth]{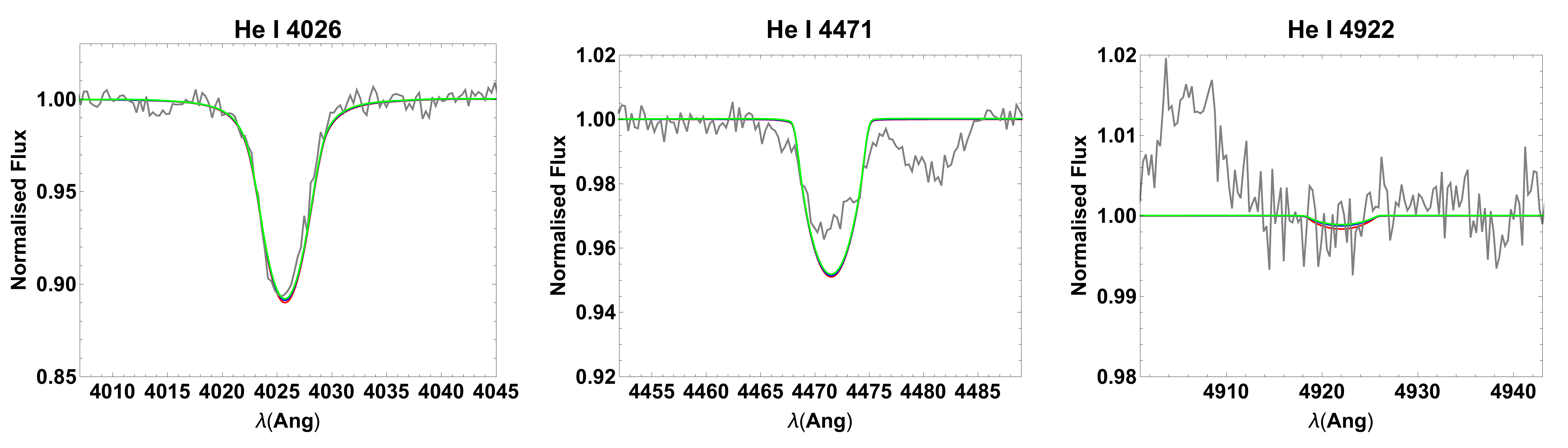}
			\caption{Resulting FASTWIND spectra for $\zeta$-Puppis with $T_\text{eff}=40$ kK, $\log g=3.64$, $R_*/R_\odot=18.6$ and $\dot M=5.2\times10^{-6}$ $M_\odot$ yr$^{-1}$. Clumping factors are $f_\text{cl}=1.0$ (red, homogeneous), $f_\text{cl}=5.0$ (blue) and $f_\text{cl}=9.0$ (green).}
			\label{comp000809}
		\end{figure*}
		\begin{figure*}[htbp]
			\centering
			\includegraphics[width=\linewidth]{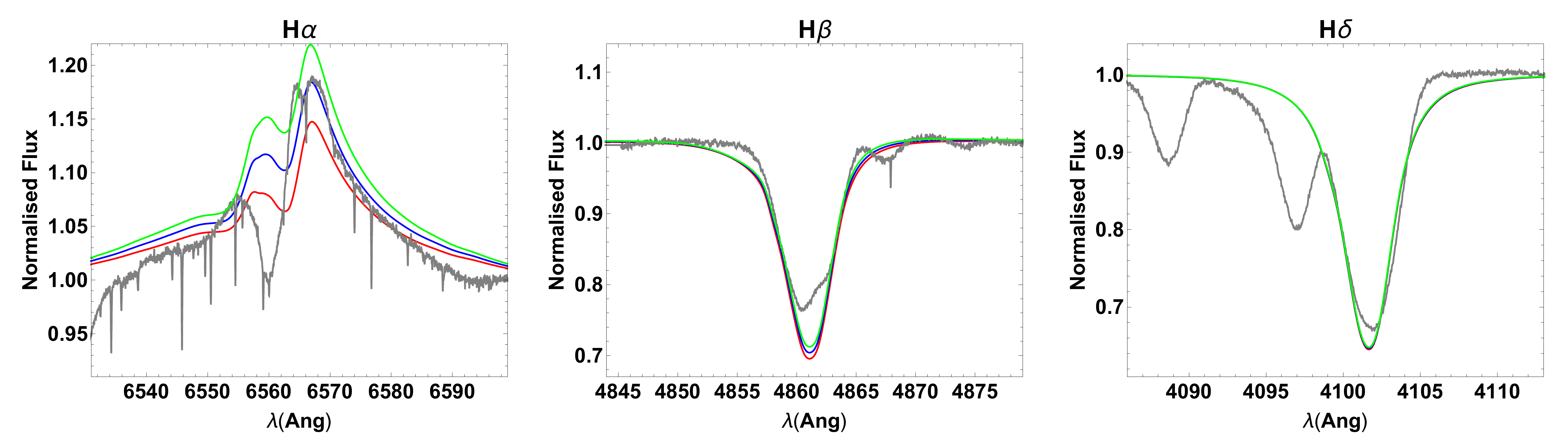}
			\includegraphics[width=\linewidth]{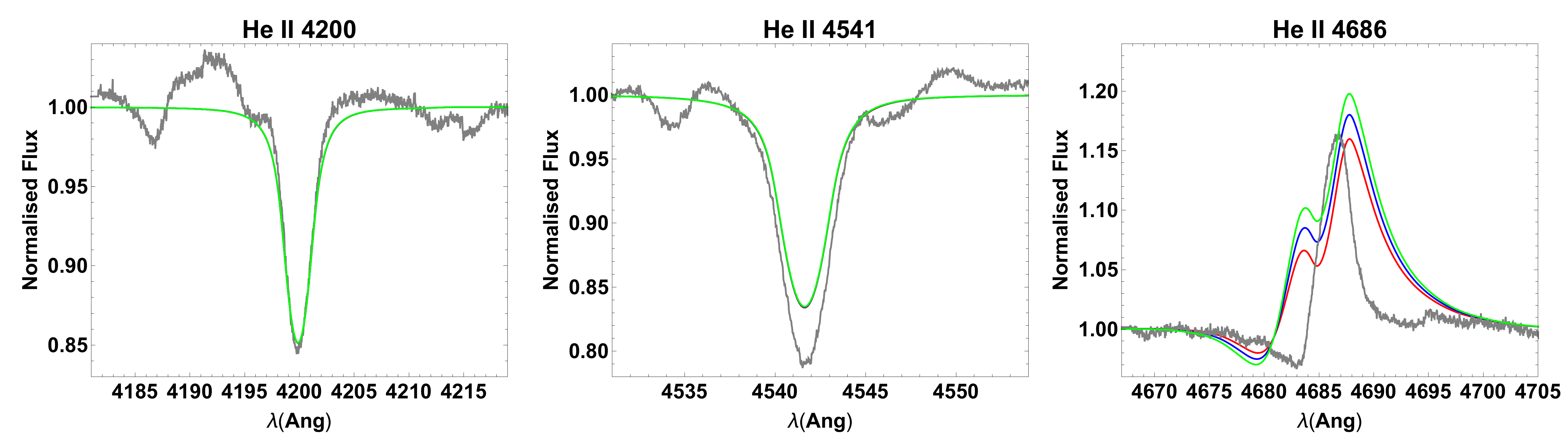}
			\includegraphics[width=\linewidth]{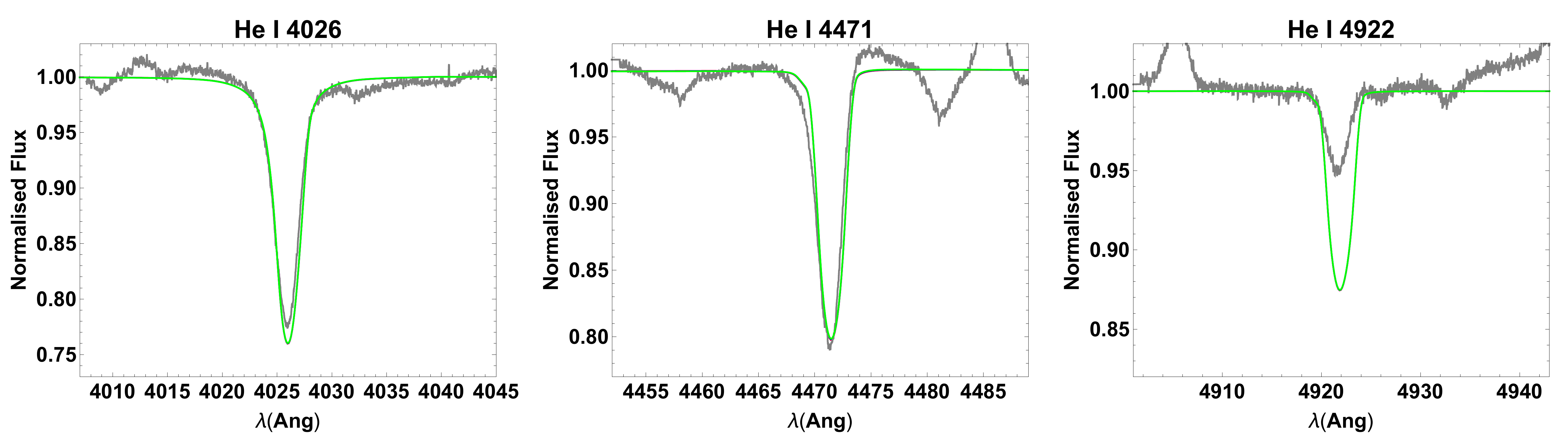}
			\caption{Resulting FASTWIND spectra for HD 163758 with $T_\text{eff}=34.5$ kK, $\log g=3.41$, $R_*/R_\odot=21.0$ \citep[see][]{bouret12} and $\dot M=3.3\times10^{-6}$ $M_\odot$ yr$^{-1}$. Clumping factors are $f_\text{cl}=5.0$ (red), $f_\text{cl}=6.0$ (blue) and $f_\text{cl}=7.0$ (green).}
			\label{comp537366}
		\end{figure*}
		\begin{figure*}[htbp]
			\centering
			\includegraphics[width=\linewidth]{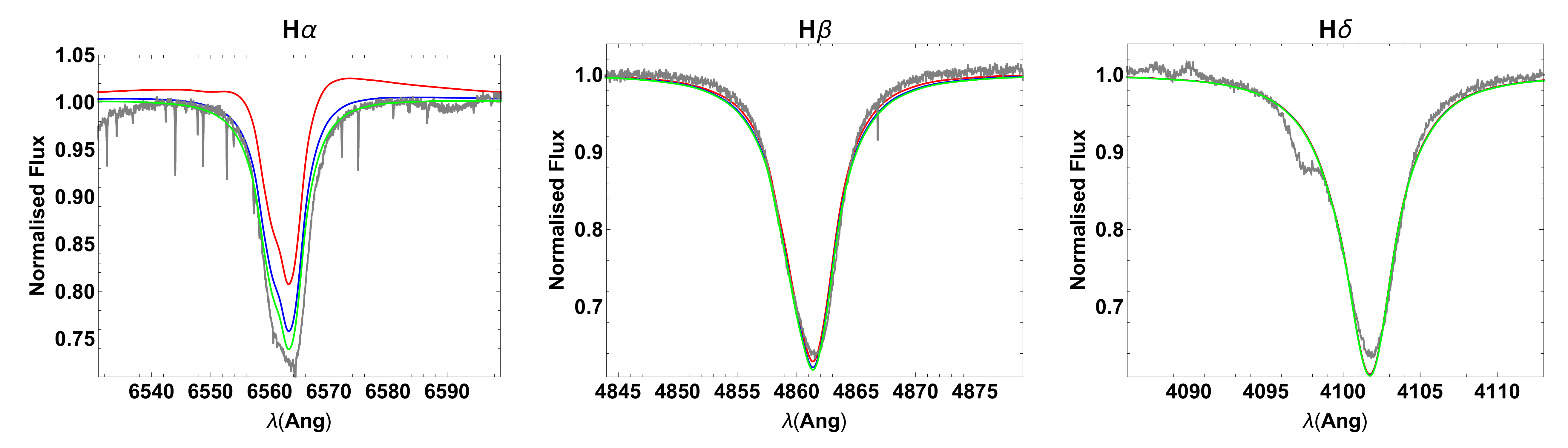}
			\includegraphics[width=\linewidth]{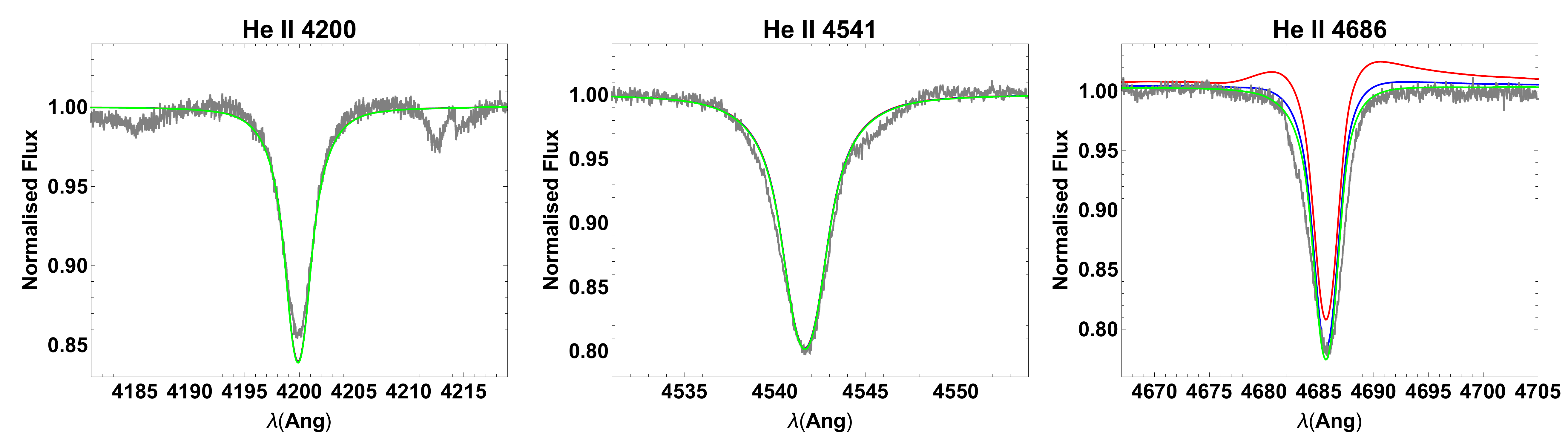}
			\includegraphics[width=\linewidth]{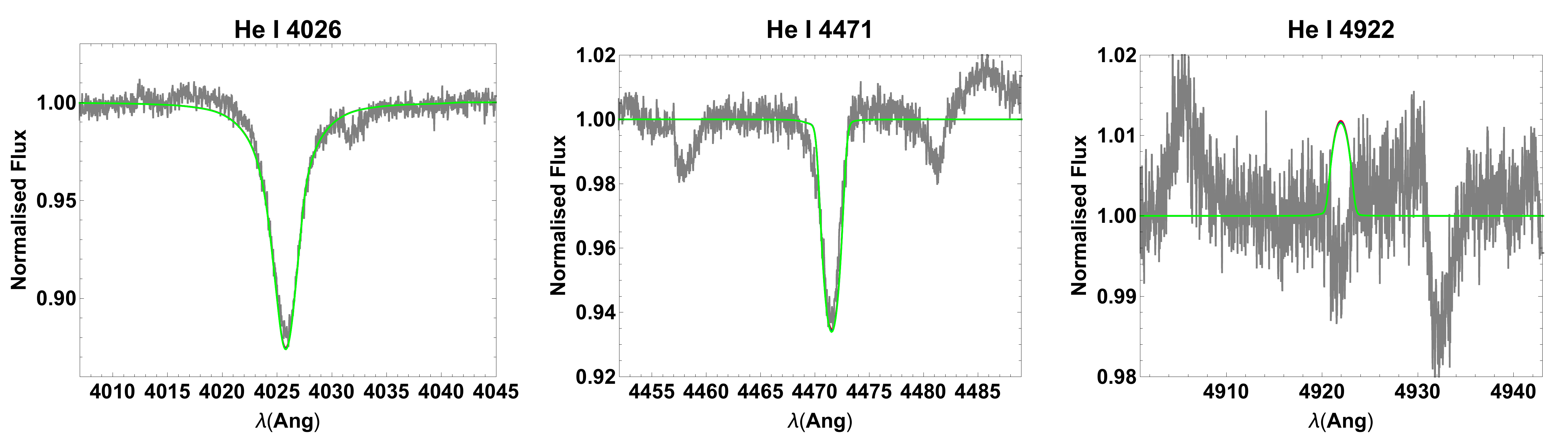}
			\caption{Resulting FASTWIND spectra for HD 164794 with $T_\text{eff}=43.8$ kK, $\log g=3.92$, $R_*/R_\odot=13.1$ \citep[stellar parameters taken from][]{krticka15} $\dot M=2.3\times10^{-6}$ $M_\odot$ yr$^{-1}$. Clumping factors are $f_\text{cl}=5.0$ (red), $f_\text{cl}=2.0$ (blue) and $f_\text{cl}=1.0$ (homogeneous, green).}
			\label{comp858687}
		\end{figure*}

\end{document}